\documentstyle[general_cite,graphicx,subfigure,graphics]{mnras}

\title[The extended 12-micron Seyferts at 8.4~GHz]{High--resolution 
radio observations of Seyfert galaxies in the extended 12-micron 
sample -- II. The properties of compact radio components.}     

\author[A.Thean, A.Pedlar, M.Kukula, S.Baum and C.O'Dea]
{Andy~Thean$^{1,2}$, Alan~Pedlar$^{2}$, Marek~J. Kukula$^{3}$, Stefi
A. Baum$^{4}$ \cr
and Christopher P. O'Dea$^{4}$\\ 
$^{1}$~Istituto di Radioastronomia del CNR, Via P. Gobetti 101,
I--40129 Bologna, Italy\\ 
$^{2}$~Jodrell Bank Observatory, University of Manchester,  Jodrell~Bank, Macclesfield,
Cheshire SK11~9DL, U.K. \\
$^{3}$~Institute for Astronomy, University of Edinburgh, Royal Observatory, Blackford Hill, Edinburgh EH9 3HJ\\
$^{4}$~Space Telescope Science Institute, 3700 San Martin Drive,
Baltimore, Maryland 21218, USA}

\date{1st Dec 2000}

\begin{document}

\maketitle

\begin{abstract}

We discuss the properties of compact nuclear radio
components in Seyfert galaxies from the 
extended 12 $\mu$m AGN sample of Rush et al. (1993). 
Our main results can be summarised as follows.

Type 1 and type 2 Seyferts produce compact radio components which are
indistinguishable in strength and aspect, indicating that their central
engines are alike as proposed by the unification model.
Infrared {\it IRAS} fluxes are more closely correlated with 
low--resolution radio
fluxes than high--resolution radio fluxes, suggesting that they are
dominated by kiloparsec--scale, extra--nuclear emission regions; 
extra--nuclear emission may be stronger in type 2 Seyferts.
Early--type Seyfert galaxies tend to have stronger nuclear 
radio emission than late--type Seyfert galaxies. 
V--shaped extended emission--line regions, indicative of `ionisation cones',
are usually found in sources with large, collimated radio outflows.
Hidden broad lines are most likely to be found in sources with 
powerful nuclear radio sources.
Type 1 and type 2 Seyferts selected by their {\it IRAS}
12 $\mu$m flux densities have well matched properties.

\end{abstract}


\section{Introduction}

Much of the progress in understanding 
the physics of active galaxies has been made by
studying the sub--class known as Seyfert galaxies.
Seyfert galaxies are more common, in terms of their space density,
than radio galaxies or quasars, and nearby examples can be identified
and studied in detail. 
Their nuclei are bright enough to be observed in most regions of the
electromagnetic spectrum using various techniques and their stellar
populations can be studied more easily than those of 
more distant classes of active galaxy.
The wide variety of good quality survey information available about
Seyfert galaxies means that valuable insights into their behaviour can
be obtained using statistical techniques i.e. samples of Seyfert
galaxies can be selected using a choice of parameters and
used to address a range of open questions.
In this paper we investigate the properties of the compact radio
components found at Seyfert nuclei by using one of the largest
homogeneously--selected samples of Seyfert galaxies available.

The central engines of Active Galactic Nuclei (AGN) are thought to 
contain supermassive black holes surrounded by energy--emitting 
accretion discs.
Unique information about Seyfert central engines can be obtained from
high--resolution radio observations.
The strengths of the radio cores show the level of nuclear activity, 
even in dust--enshrouded sources, and may
indicate the masses of the supermassive black holes 
\cite{Franceschini98}.
In addition, small--scale, collimated radio structures reveal outflows
which are thought to dominate the
structure and dynamics of Seyfert narrow--line regions
(\pcite{Falcke98}; \pcite{Axon98}; \pcite{Capetti99})
and probably indicate the position angles of the
central accretion discs.

Despite the influence of radio outflows on the narrow--line region, 
at present they do not play an important r\^{o}le in Seyfert 
unification models.
According to these models there is one
population of intrinsically similar objects whose orientation 
determines their observed properties. 
It is proposed that the sub--parsec--scale   
broad--line--emitting region in type 2 sources is hidden by a 
suitably oriented dusty torus and only narrow emission lines from
the larger narrow--line--emitting region are seen; in type 1 sources the
orientation of the torus allows both broad and narrow lines to be
observed (see \pcite{AntonucciReview}).
 
Recent radio surveys indicate that the radio properties of type 1 and
type 2 Seyferts are generally well matched, in support of the
unification model (\pcite{Edelson87};  
\pcite{Ulvestad+W89}; 
\pcite{Giuricin90}; 
\pcite{Kukula95}; 
\pcite{RushM+E96};
\pcite{Nagar99};
\pcite{Morganti99}).
Early radio surveys which found differences between type 1 and type 2  
Seyferts 
(\pcite{Ulvestad+W84a}; 
\pcite{Ulvestad+W84b})
were probably affected by selection 
effects (\pcite{Salzer89}; \pcite{Wilson91}),
but there remain a handful of recent results which 
are not readily explained by the unification model e.g. 
\scite{Roy94} 
found that the detection rate of compact radio cores smaller than 0.1 arcsec 
was significantly lower for
type 1 Seyferts than for type 2 Seyferts, and observations by
\scite{Kukula95} show that an excessively high fraction of 
type 1 Seyferts are unresolved at 0.25 arcsec resolution.

This paper is organised as follows;
in Section \ref{sample.sec} we describe the sample, the data and the
statistical tests used in our analysis, 
in Section \ref{compact.sec} we compare the compact nuclear radio components 
found in type 1 and type 2 Seyfert galaxies,
in Section \ref{IR.sec} we discuss the origin of the 
{\it IRAS} emission from Seyfert galaxies,
in Section \ref{host.sec} we examine the relationship between host galaxy
and the central engine,
in Section \ref{Cone.sec} we examine the radio properties of Seyferts with
V--shaped extended emission line regions,
in Section \ref{HBLR.sec} we examine the radio properties of Seyferts with
known hidden broad--line regions,
in Section \ref{bias.sec} we discuss the sample selection
and in Section \ref{summary.sec} we give a summary of our results.

A value of H$_{\circ}$ = 75 km$\,$s$^{-1}$Mpc$^{-1}$ is assumed throughout.

\section{SAMPLE, DATA AND STATISTICAL TESTS}
\label{sample.sec}

Our results are based on radio observations of the extended 12
$\mu$m sample of \scite{RMS93} made with the VLA in A--configuration at
8.4 GHz \cite{Thean99}.
These 0.25--arcsec--resolution observations allow elongated radio
structures tens of parsecs in size to be resolved and enable 
radio components smaller than 3.5 arcsec to be isolated from
kiloparsec--scale, low--brightness--temperature emission
(see \pcite{TheanNFRA} for a discussion of instrumental limitations).
The typical 1--$\sigma$ noise level was 50 $\mu$Jy/beam.
In addition to our new observations we make use of 
archive data from the {\it IRAS}
survey as obtained from \scite{RMS93}, the 1.4 GHz NRAO
VLA Sky Survey (NVSS) described by \scite{Condon98}, the Lyon--Meudon
Extragalactic Database (LEDA) described by \scite{Paturel96} and the
NASA/IPAC Extragalactic Database (NED) described by \scite{NED}. 
All archive data used in our analysis are given in Appendix \ref{archive.sec}. 

The extended 12 $\mu$m AGN sample 
is an extension of the original 12 $\mu$m AGN sample of
\scite{SM89} to fainter flux levels using the 
{\it IRAS Faint Source Catalogue Version
2} \cite{Moshir91}.
From an initial sample of 893 mid--infrared--bright sources,
AGN catalogues were used to define a 
sub--sample of active galaxies which contains 118 objects, the majority of
which are Seyfert galaxies.
The sample is one of the largest homogeneously--selected samples of Seyferts
available and appears to contain representative populations of
galaxies with type 1 and type 2 optical spectra; \scite{SM89} proposed
that the {\it IRAS} 
12 $\mu$m waveband carries an approximately constant fraction, around 20\%,
of the bolometric flux for quasars and both types of Seyfert 
(selection effects are discussed in Section \ref{bias.sec}).
In order to restrict our analysis to well--identified
Seyfert galaxies, we have excluded 11 sources from our analysis of 
the extended 12 $\mu$m AGN sample and reclassified two. 
The exclusion of two LINERs (NGC 1097 and NGC 3031) and
3 starburst galaxies (NGC 1386, Arp 220 and NGC 6810) and the
re-classification  
of two Seyfert 1s as Seyfert 2s (Markarian 6 and NGC 2992) follows
\scite{Dopita98}, the exclusion of one starburst galaxy (NGC 34) follows
\scite{Mulchaey96} and the exclusion of five radio--loud sources (3C 120,
3C 234, 3C 273, 3C 445 and OJ 287) follows Paper I \cite{Thean99}.
The resulting sample contains 47 type 1 Seyferts and 60 type 2 Seyferts.
As pointed out by \scite{Hunt99I}, 
the accurate spectroscopic classification of the sample is still not
complete.  
In Section \ref{ID.sec} we identify 38 sources which are
classified as normal or high--far--infrared galaxies by \scite{RMS93},
but which are classified as Seyfert galaxies in the NED.

Our analysis has been carried out using univariate
two--sample tests and bivariate correlation tests based on 
Survival Analysis techniques, which allow information about
censored data (size and flux density upper--limits) to be used efficiently.
Tests based on Survival Analysis were implemented using the 
software {\small ASURV} \cite{Lavalley92}.
Two--sample tests available with {\small ASURV} have been discussed by
\scite{Feigelson+N85} and include two versions of the Gehan's test 
(permutation variance and hypergeometric variance), the Logrank 
test, the Peto \& Peto test and the Peto \& Prentice test.
Correlation tests have been discussed by \scite{Isobe86} 
and include the Cox Proportional Hazard test,
Kendall's Tau and Spearman's Rho.
For ease of comparison, we have maintained the use of these tests 
even when no censored data have been used,
in these cases the Peto \& Prentice test reduces to the Gehan's test
and we are able to apply the Kolmogorov--Smirnov (KS) test.
We have used these tests to determine the probability of obtaining a 
test statistic under a given null hypothesis; 
usually the null hypothesis
states that two distributions are drawn at random from the same parent
population for two--sample tests, or that there is no correlation 
between two variables for correlation tests. 
We adopt the convention that a result is 
{\it marginally significant}
if the probability of the null hypothesis, P(null), 
is less than 5\% for at least one test,
{\it significant} 
if  P(null) is less than 5\% for all tests,
{\it highly significant} 
if P(null) is less than 1\% for all tests and 
{\it extremely significant} 
if P(null) is less than 0.1\% for all tests. 
Where censored data has been used, cumulative distributions are
plotted using the Kaplan--Meier estimator which is a non--parametric, 
maximum--likelihood estimator of a randomly--censored sample 
(see \pcite{Feigelson+N85}). 
For each plot we display the mean uncertainty of all points on the
cumulative distribution as calculated using {\small ASURV}.

\section{Compact radio components in type 1 and type 2 Seyferts}
\label{compact.sec}

\subsection{The strength of the radio emission}
\label{S8.4.sec}

\begin{figure} 
\centerline{
       \includegraphics[angle=0,width=8.5cm]{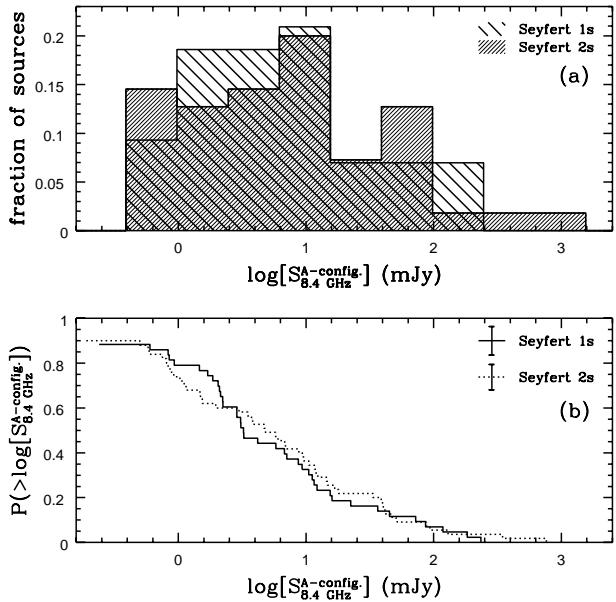}}
\caption{(a) Histograms showing the fractional 
8.4 GHz A--configuration flux density distributions of the 38 type 1 and 48 type 2
Seyferts from the extended 12 $\mu$m sample which were detected.   
(b) The cumulative flux density distributions of the 43 type 1 and 55 type 2
Seyferts observed, as given by the Kaplan--Meier
estimator where the y axis gives the  
probability that a source is brighter than a given flux density.  
The flux density distributions of the two Seyfert types are 
statistically indistinguishable.}
\label{Aflux.fig}
\end{figure}

\begin{figure} 
\centerline{
       \includegraphics[angle=0,width=8.5cm,clip,trim=20 0 230 0]{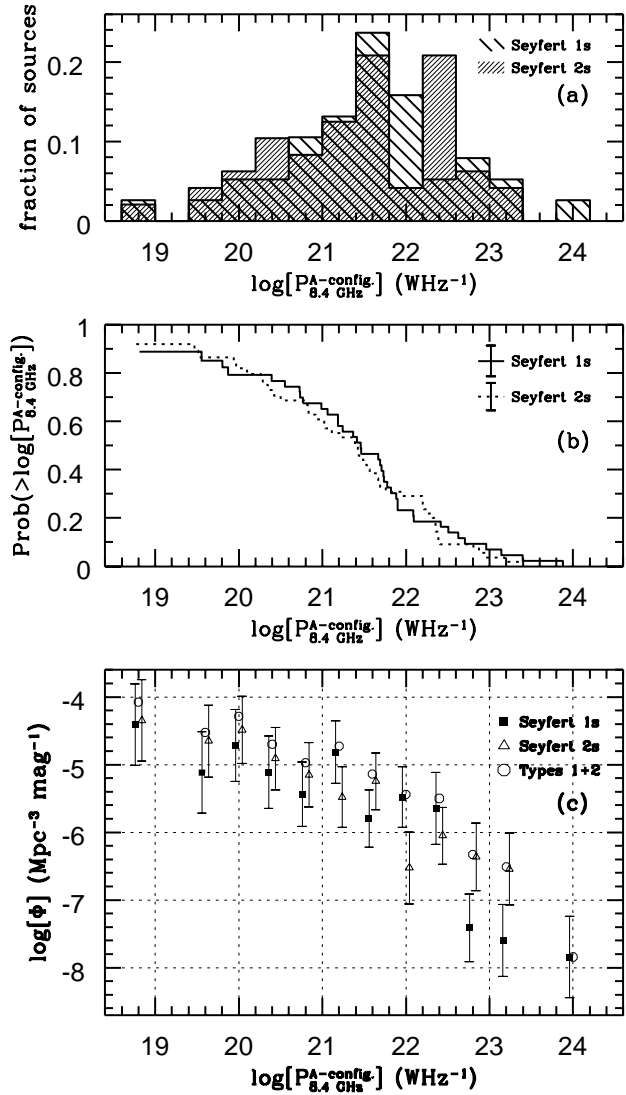}
}
\caption{(a) The fractional 8.4 GHz A--configuration luminosity 
distributions of the 38 type 1 and 48 type 2 Seyferts
from the extended 12 $\mu$m sample which were detected.
(b) The cumulative 8.4 GHz A--configuration 
luminosity distributions of the 43 type 1 and 55 type 2 Seyferts observed,
as given by the Kaplan--Meier estimator.
The luminosity distributions of the two Seyfert types are 
indistinguishable and this suggests that they contain the same type
of central engine.
(c) The absolute space densities of the two Seyfert types have been
estimated by normalising the differential 8.4 GHz VLA A-configuration radio
luminosity functions.
For display purposes, points representing the bin centres of Seyfert type 2s
and Seyfert types 1 and 2, have been offset by values of log(P) = -0.04
and log(P) = 0.04 respectively. 
}
\label{Alum.fig}
\end{figure}

Figure \ref{Aflux.fig} 
shows the 8.4 GHz A--configuration flux density
distributions of type 1 and type 2 Seyferts from the extended 12 $\mu$m sample.
There is no significant difference between the 
8.4 GHz A--configuration flux densities of type 1 and type 2 Seyferts. 
The probability of the null hypothesis, that the two flux density
distributions are drawn at random from the same parent population, is: 
89.98\% (Gehan's, perm.), 90.02\% (Gehan's, hyper.), 92.95\% 
(Logrank test), 91.21\% (Peto \& Peto) and 91.19\% (Peto \& Prentice).  

Figures \ref{Alum.fig}(a) and \ref{Alum.fig}(b) show the 8.4 GHz 
A--configuration luminosity distributions of type 1 and type 2 
Seyferts from the sample.
The two distributions are statistically indistinguishable, and span 
approximately the same range in luminosity, from 1$\times$10$^{18.6}$
to 1$\times$10$^{24.0}$ WHz$^{-1}$.  
The probability of the null hypothesis, that the two luminosity
distributions are drawn at random from the same parent population, is: 
70.86\% (Gehan's, perm.), 70.87\% (Gehan's, hyper.), 77.76\% (Logrank
test), 71.22\% (Peto \& Peto) and 71.48\% (Peto \& Prentice).  
In order to estimate the magnitude of the systematic difference in luminosity 
which would allow the null hypothesis to be rejected we have made 
trials with the observational data.
By scaling the luminosity of the Seyfert 1 sub--population and repeating 
the two--sample tests, we find that
the probability of the null hypothesis is less than 5\% for all 
statistical tests when scaling factors less than or equal to one 
sixth or greater than or equal to four are applied (see Appendix
\ref{tests.sec} for test results using various scaling factors).

Figure \ref{Alum.fig}(c) shows the differential 8.4 GHz
A--configuration luminosity function of the 
extended 12 $\mu$m Seyfert sample (the construction of the  differential 
luminosity function is described in Appendix \ref{difflum.sec}). 
According to the unification model, the relative space densities of
Seyfert 1s  and Seyfert 2s are determined by the typical opening angle of the
circum--nuclear torus.
We have used 3 methods to estimate a typical opening angle from the 
differential radio luminosity function, all methods give a type 1 to 
type 2 ratio less than unity, but  
this quantity is poorly constrained.
Summing $\Phi$(L) over bins containing both Seyfert type 1s and Seyfert 
type 2s yields a cone opening angle of 110$^{+20}_{-50}$ degrees (using all such
bins) and  150$^{+10}_{-50}$ degrees (using only those 3 bins containing more
than 4 sources); errors are calculated by taking the uncertainty in
$\Phi$(L) values as proportional to the Poisson noise in each bin.
Integrating linear power law fits over the observed range in radio power 
(1$\times$10$^{18.4}$ to 1$\times$10$^{24.0}$ WHz$^{-1}$) 
yields an opening angle of 60$^{+10}_{-10}$ degrees, but because the type 1
and type 2 fits have different gradients this value is sensitive to the lower 
radio power limit used (fits were sought only
for luminosity bins containing more than 4 sources and the
uncertainty in the opening angle is derived from the uncertainty 
in the fits).

\subsection{Radio source sizes}
\label{D.sec}

\begin{figure} 
\centerline{
       \includegraphics[angle=0,width=8.5cm]{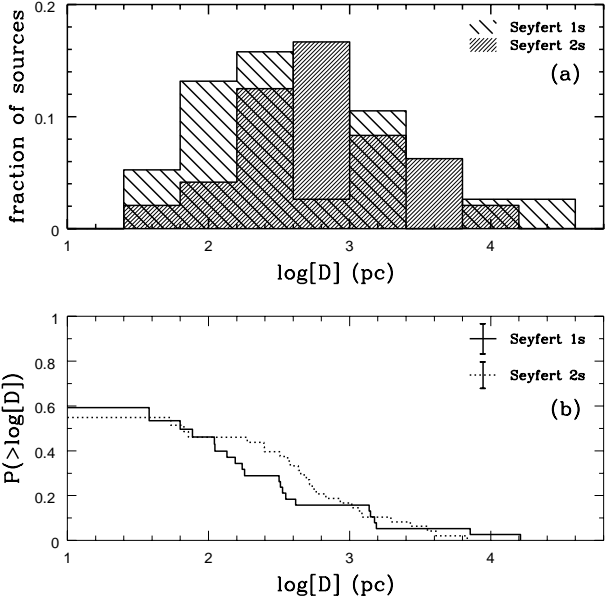}
}
\caption{(a) The fractional size distributions of the 20 type 1 and 25 type 2
Seyferts from the extended 12 $\mu$m sample which were resolved. 
(b) The cumulative size distributions of the 
38 type 1 and 48 type 2 Seyferts detected, 
as given by the Kaplan--Meier estimator
where the y axis gives the probability that a source is
larger than a given size. 
The two size distributions are statistically indistinguishable
indicating that type 1 and type 2 Seyferts have an equal capacity for
producing extended nuclear radio components.}
\label{size.fig}
\end{figure}

\begin{table}
\tiny
\begin{center}
\begin{tabular}{|l|c|c|c|c|c|} \hline 
{\bf Radio} & & \multicolumn{2}{c|}{\bf Full sample} & \multicolumn{2}{c|}{\bf
S$\rm _{8.4}>$ 5 mJy} \\ 

{\bf components} & {\bf T$\rm _{\rm rad}$} & {\bf Type 1} & {\bf
Type 2} & {\bf Type 1} & {\bf Type 2} \\ 
\hline
1 unresolved          & U & 19~(40\%) &  22~(37\%) & ~6~(32\%) & ~8~(30\%) \\
1 slightly--          & S & ~9~(19\%) &  ~6~(10\%) & ~7~(37\%) & ~4~(15\%) \\
resolved              & & &  & & \\
2                     & L & ~3~~(6\%) &  ~6~(10\%) & ~0~~(0\%) & ~4~(15\%)\\
$>$2 linearly--       & L & ~4~~(9\%) &  ~7~(12\%) & ~3~(16\%) & ~6~(22\%) \\
aligned               &  &  &  &  & \\
Diffuse               & D & ~0~~(0\%) &  ~2~~(3\%) & ~0~~(0\%) & ~2~~(7\%) \\
Diffuse +             & D+U & ~0~~(0\%) &  ~1~~(2\%) & ~0~~(0\%) & ~1~~(4\%) \\
unresolved            &  &  &  &  &  \\
Ambiguous             & A & ~3~~(6\%) &  ~3~~(5\%) & ~3~(16\%) & ~2~~(7\%) \\
unobserved            & - & ~4~~(9\%) &  ~5~~(8\%) & - & - \\
undetected            & - & ~5~(11\%) &  ~8~(13\%) & - & - \\
\hline
Total           &   & 47~~~~~~~~  &  60~~~~~~~~ & 19~~~~~~~~ & 27~~~~~~~~ \\
\hline
\end{tabular}
\end{center}
\caption{The 8.4 GHz A--configuration radio structures of Seyferts from the
extended 12 $\mu$m sample. The true distribution of radio morphologies
is more accurately represented by a sub--sample of sources
brighter than 5 mJy/beam for which high--dynamic--range maps are available.}
\label{struct.tab}
\end{table}

Table \ref{struct.tab} shows the number of sources of each structural
type within the full sample and within a sub--sample of sources
with 8.4 GHz flux densities greater than 5 mJy/beam (typically 100--$\sigma$). 
Structural types, T$_{\rm rad}$, follow the notation used by 
\scite{Ulvestad+W84a}: U for single
unresolved sources, S for single slightly--resolved sources, A for 
sources with ambiguous structures, D for sources with diffuse structures
and L for sources with possible linear structures.

Figure \ref{size.fig} shows the size distributions 
of the nuclear radio structures in type 1 and type 2 Seyferts from the 
extended 12 $\mu$m sample.
Each Seyfert type shows an approximately equal fraction of 
unresolved sources (around half) and the two size 
distributions are statistically indistinguishable. 
The probability of the null hypothesis, that type 1 and type 2 
size distributions are drawn at random from the same parent population,
is: 53.00\% (Gehan's, perm.), 53.16\% (Gehan's, hyper.), 78.93\%
(Logrank), 59.69\% (Peto \& Peto) and 59.51\% (Peto \& Prentice).
To estimate the magnitude of the systematic difference in sizes
which would allow the null hypothesis to be rejected we have made
trials with the observational data.  
By scaling the sizes of the Seyfert 1 sub--population 
and repeating the two--sample tests, we find that 
the probability of the null hypothesis is less than 5\% for all 
statistical tests when scaling factors less than or equal to one 
sixth and greater than or equal to 10 are applied (see Appendix
\ref{tests.sec} for test results using various scaling factors).

\begin{figure} 
\centerline{
       \includegraphics[angle=0,width=8.5cm,clip, trim=0 0 0 200]{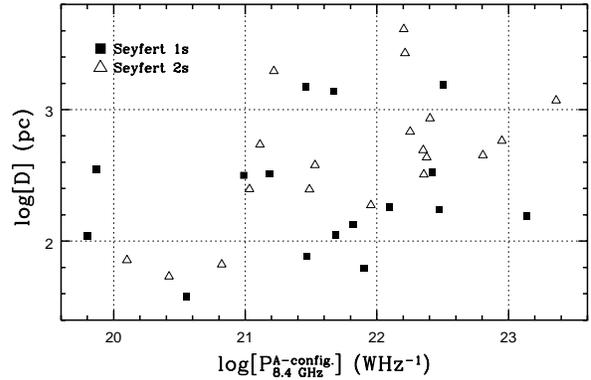}}
\caption{Radio size is plotted against radio power 
for collimated sources (types S and L) from the extended 12 $\mu$m
Seyfert sample.
There is a statistically significant correlation between radio power
and radio size for type 2 Seyferts, but not for type 1 Seyferts.}
\label{size-lum.fig}
\end{figure}

Figure \ref{size-lum.fig} shows how radio size varies with  
radio power for sources which have slightly--resolved, double or
linearly--aligned radio components (structural types S or L).
There is a statistically significant correlation between radio size and radio
power for this sub--population. 
The probability of the null hypothesis, that there is no correlation
between radio size and radio power, is:
2.19\% (Cox proportional Hazard), 1.91\%
(Kendall's Tau) and 1.37\% (Spearman's Rho).
Interestingly, the correlation is only significant for type 2 sources.
For the 16 type 1 sources the probability of the null hypothesis is:
50.43\% (Cox proportional Hazard), 52.85\%
(Kendall's Tau) and 56.90\% (Spearman's Rho).
For the 19 type 2 sources the probability of the null hypothesis is: 
3.04\% (Cox proportional Hazard), 1.91\%
(Kendall's Tau) and 2.02\% (Spearman's Rho).

For the extended 12 $\mu$m Seyfert 
sample as a whole, sources with type S or type L
radio sources have higher radio luminosities than Seyferts of other 
structural types, but this result may be partly attributable 
to the difficulty in detecting collimated structures in faint sources;
resolved sources can be wrongly classified as unresolved if their secondary
flux components fall below the flux limit.
For the Seyfert sample as a whole, the probability of the null hypothesis,
that sources with type S or type L radio structures and sources with
other types of radio structure have luminosity distributions which 
are drawn at random from the same parent population, is:
0.18\% (Gehan's, perm.), 0.11\% (Gehan's, hyper.), 0.02\%
(Logrank), 0.16\% (Peto \& Peto) and 0.18\% (Peto \& Prentice).
However, for a bright sub--population of 46 sources with 8.4 GHz
A--configuration flux densities higher than 5 mJy/beam there is no significant
difference between the radio luminosities of sources with type S or
type L radio structures and sources with other structural types, the
probability of the null hypothesis is:
45.47\% (Gehan's, perm.), 45.34\% (Gehan's, hyper.), 96.39\%
(Logrank) and 96.39\% (Peto \& Peto).

Note that some of the most luminous sources in the sample are unresolved 
at our resolution.
This sub--population of bright, unresolved sources includes
two of the three most luminous radio sources in the sample (Markarian 231
and Markarian 348) and the three most `radio--loud' Seyferts in the
sample (Markarian 348, F01475-0704 and NGC 7213 have the highest 
8.4 GHz A--configuration to {\it IRAS} 60 $\mu$m flux ratios).

\subsection{Implications: the central engines 
of type 1 and type 2 sources are alike.}
\label{compactImp.sec}

Our results, which are based on the largest sub--arcsecond--resolution
radio imaging survey of a homogeneously--selected sample of Seyfert
galaxies available, provide the most stringent comparison 
of the radio powers and radio sizes of Seyfert nuclei to date.
The fact that we find no significant differences between the
strengths, sizes and morphologies of compact radio components in 
type 1 and type 2 Seyfert galaxies is evidence that their central
engines are alike, as proposed by unification models.

According to the unification model, obscuration by a 
circum--nuclear dusty tori gives rise to the differences between the 
observational properties of type 1 and type 2 Seyferts, but, 
since dust is transparent at radio wavelengths, their radio
luminosities are expected to be equivalent.
In agreement with previous high--resolution 
surveys (\pcite{Kukula95}; \pcite{Nagar99}),
we have found that the 8.4 GHz A--configuration 
radio luminosities of the two Seyfert types are
statistically indistinguishable (\S \ref{S8.4.sec}) and
we estimate that if systematic differences in luminosity are present
they are less than a factor of six. 
We find no evidence of orientation--dependent free--free absorption, 
but our luminosity comparison is not very sensitive to such effects because
we have included extra--nuclear flux components and because 
free--free absorption is not expected to be strong at 8.4 GHz
(e.g. \pcite{Krolik89}). 

The fact that the radio structures of type 1 and type 2 Seyferts 
are statistically indistinguishable (\S \ref{D.sec}) 
is also consistent with the Seyfert unification model.
According to the unification model, the two Seyfert types have
an equal capacity for producing collimated radio structures
and differ only by their orientation.
Orientation is expected to make the observed radio structures in
Seyfert type 2s systematically larger than those in Seyfert type 1s
(if the axis of the obscuring torus is aligned with the radio
collimation axis, type 1 Seyferts should be geometrically
fore--shortened), but variations in intrinsic source length are
expected to dominate the effects of orientation, especially
for models of tori with intermediate opening angles.
We estimate that our tests are only able to reject the 
hypothesis that the sizes of the two Seyfert types are equivalent
for systematic differences in size greater than an approximate factor
of 8 and therefore they are probably insensitive to the 
effects of orientation. 
The statistically insignificant trend for the radio sources in 
type 2 Seyferts to be larger than those in Seyfert 1s 
(Fig. \ref{size.fig}) and the relative fractions of sources
with type S and type L radio structures in a bright sub--sample
(Table \ref{struct.tab}) are in the sense predicted by the unification
model. 

Observations of the Early-type Seyfert sample \cite{Nagar99}, 
and the CfA Seyfert sample \cite{Kukula95}
have shown that type 1 Seyferts are more likely to 
have unresolved nuclear radio structures than type 2 Seyferts. 
For the  CfA Seyfert sample and a volume--limited southern Seyfert sample
\cite{Morganti99} the 
difference between the size distributions of type 1 and type 2 
Seyferts is statistically significant
(this is not the case for the Early-type Seyfert sample).
VLA A--configuration observations of the CfA sample at 8.4 GHz 
show that only one type 1 Seyfert is fully resolved
compared to nine type 2 Seyferts.
The probability that the radio size distributions of type 1 and 
type 2 CfA Seyferts are drawn at random from the same parent population is: 
0.59\% (Gehan's, perm.), 0.53\% (Gehan's, hyper.), 1.28\%
(Logrank), 0.84\% (Peto \& Peto) and 0.78\% (Peto \& Prentice).
Note however, that the difference in size between type 1 and type 2 Seyferts
in the CfA sample may be related to a bias against 
distant type 2 Seyferts (see \S \ref{12umFlux.sec}); whether a similar
bias affects the southern Seyfert sample of Morganti et al. is unclear.
A bias against distant type 2 Seyferts can affect observed size
distributions because a resolved radio source is more likely to appear
unresolved if it is more distant; 
for most sources, this is because secondary flux components fall 
below the flux limit rather than the effect of fixed angular resolution.
If the redshifts of all type 2 Seyferts in the CfA sample are scaled
by a factor of 1.6 (the average difference between the type 2 and 
type 1 redshifts, as estimated from linear fits to the 
cumulative redshift distributions), three
previously resolved sources become unresolved and the
size difference between type 1 and type 2 Seyferts is no longer 
statistically significant. 

The only difference we have found between the compact radio components
in type 1 and type 2 Seyferts is that the latter show a significant 
correlation between radio size and radio power whereas the former do not.
Weaker evidence of this difference has been found previously
\cite{Nagar99}.
The difference is in the sense predicted by the Seyfert unification model, 
according to which a correlation between radio size and
radio power should be clearest for type 2 Seyferts;
the observed angular sizes of elongated sources whose axes are close to the
line of sight are less reliable indicators of true source sizes than 
the angular sizes of sources lying close to the plane of the sky 
i.e. it is possible that large type 1 sources can project 
small apparent sizes whereas large type 2 sources cannot.
According to the unification model, the 
radio sizes of type 1 Seyferts should show a looser,
flatter dependence on radio power than those of type 2 Seyferts. 
Our data are broadly consistent with such predictions, but
more data is required to test them.
The correlation between radio size and radio power for type 2
Seyferts confirms the results of previous
studies (\pcite{Ulvestad+W84b}; \pcite{Giuricin90}; \pcite{Nagar99}; 
\pcite{Morganti99}).

The size of an expanding outflow is a primary indicator of its
age and therefore 
one interpretation of the matched size distributions of
type 1 and type 2 Seyferts is that they evolve similarly and have
similar ages. 
Recent HST spectroscopy has been used to provide an upper--limit of
1.5$\times$10$^{5}$ years on the age of the kiloparsec radio outflow
in Markarian 3 \cite{Capetti99}. 
Our results confirm that compact nuclear outflows on this scale are
rare, and suggest that around
80\% to 90\% of Seyfert nuclear outflows are younger than that of
Markarian 3. 
Compact nuclear outflows appear to be
small or absent in a considerable fraction of sources; 
approximately 30\% of sources for which high--dynamic--range
maps are available are unresolved (smaller than around 100 parsecs).
This fraction is too high to be explained by models in which all
Seyferts produce radio outflows which advance at the speed of the
largest sources and probably reflects a combination of 
factors such as instrumental limitations, 
the fraction of Seyferts capable producing radio 
outflows, duty cycle patterns and outflow
evolution. 
Note that our observations are insensitive to
low--brightness--temperature flux components and therefore 
source sizes have been under--estimated in cases where the nucleus
produces extended, diffuse emission.

We can be confident that those sources with the youngest radio outflows are to
be found in the sub--population of unresolved sources. 
Possible examples of young outflows are provided by Markarian 231 and
Markarian 348, which share some of the properties of
Compact Symmetric Objects (CSOs) and Gigahertz Peaked Spectra (GPS) galaxies 
\cite{Ulvestad99} which are probably young radio--loud AGN.
If the compact radio sources in Markarian 231 and Markarian 348 
are indeed caused by young nuclear outflows, the fact that they are more
powerful than the largest sources in the sample implies that radio
outflows are particularly luminous in their early stages; 
similar suggestions have been made to explain the properties of 
the CSO/GPS population (see \pcite{ODea98}).
Furthermore we may speculate that if they are to fit the 
correlation between radio size and radio power shown 
in Figure \ref{size-lum.fig}, Markarian 231 and Markarian 348 will 
decrease in luminosity rapidly (a few orders of magnitude before reaching 
scales of hundreds of parsecs) and if the correlation between size and power 
represents an evolutionary sequence, their luminosities will increase 
again in later stages of evolution.  
Note that both Markarian 231 and Markarian 348 show diffuse 
radio emission tens of kiloparsecs in extent, but
whether such diffuse outflows are caused by previous AGN activity or 
starburst--driven superwinds is unclear (\pcite{Baum93};
\pcite{Colbert96II}; \pcite{Colbert98III}).

Finally, we note that
\scite{Franceschini98} have shown a correlation between
the 5 GHz radio power of a galactic nucleus and the dynamical 
mass of its central black hole.
If this result can be generalised,
our results imply that the masses of the black holes in 
type 1 and type 2 Seyferts are equivalent.
Given the steep dependence of radio power on black hole mass claimed by
\scite{Franceschini98}, the radio luminosities we have observed imply a range
of black hole masses spanning only 2 orders of magnitude (around 10$^{7.5}$
to 10$^{9.5}$ M$_{\odot}$ for a radio spectral index of $-$0.7).

\section{THE ORIGIN OF IRAS EMISSION FROM SEYFERT GALAXIES}
\label{IR.sec}

\subsection{High--resolution radio fluxes}
\label{S8.4-IR.sec}

\begin{figure*} 
\centerline{ \includegraphics[angle=0,width=17.5cm]{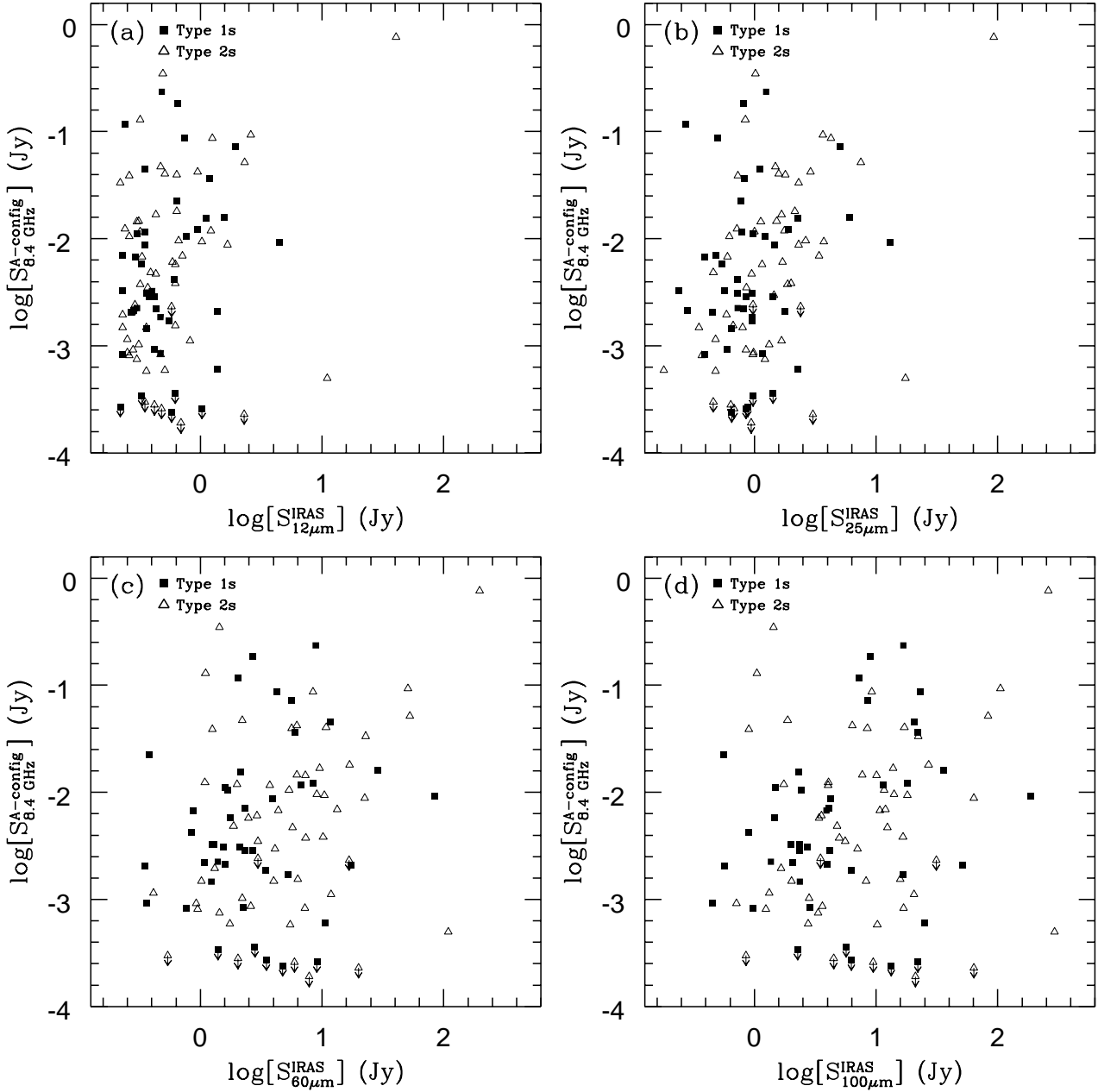}}
\caption{VLA 8.4 GHz A--configuration radio flux densities are plotted against 
(a) 12 $\mu$m, (b) 25 $\mu$m, (c) 60 $\mu$m and (d) 100 $\mu$m
{\it IRAS} flux densities for Seyferts from the extended 12 $\mu$m sample. 
The only significant correlation between 8.4 GHz A--configuration fluxes
and {\it IRAS} fluxes is found at  25 $\mu$m; there are 
marginally significant correlations at 12 $\mu$m and
60 $\mu$m, but no evidence of a correlation at 100 $\mu$m.
Radio flux density upper-limits are denoted by arrows.}  
\label{IR-Sa.fig}
\end{figure*}

\begin{figure} 
\centerline{
\includegraphics[angle=0,width=8.5cm,clip,trim=160 0 0 0]{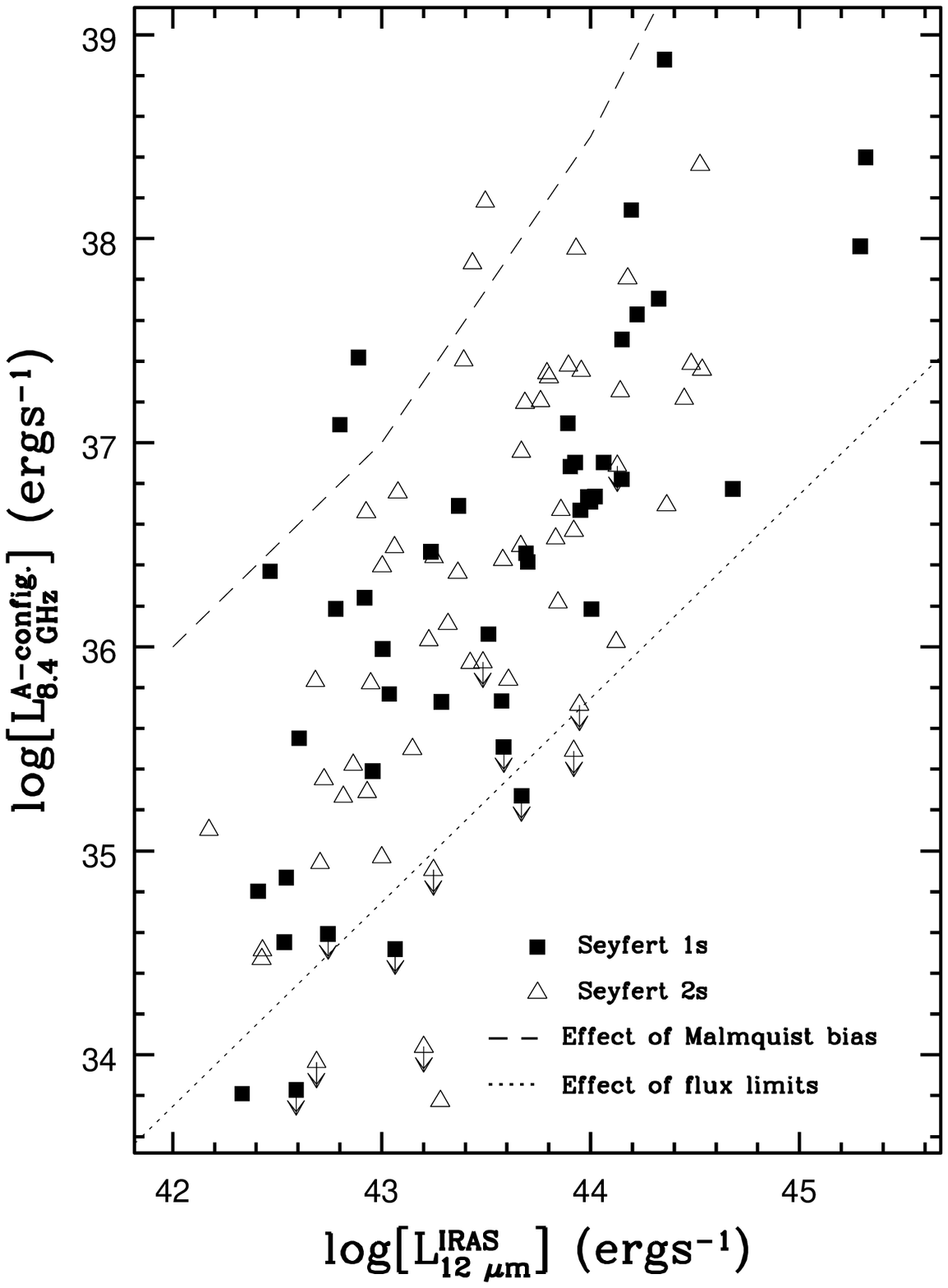}
}
\caption{8.4 GHz A--configuration luminosities are plotted
against {\it IRAS} 12 $\mu$m luminosities. 
The distribution of data points is restricted by the Malmquist bias
and the observational flux limits, the remaining 
region is evenly filled and evidence for a
correlation between the two variables is weak.}
\label{12-rad.fig}
\end{figure}

In Figure \ref{IR-Sa.fig} the  8.4 GHz A--configuration flux densities of
Seyferts from the extended 12 $\mu$m sample are plotted against 
their {\it IRAS} flux densities. 
The probability of the null hypothesis, that there is no  
correlation between 8.4 GHz A--configuration fluxes and  
{\it IRAS} fluxes for each {\it IRAS} waveband, 
is given in Table \ref{rad-IR.tab}.
The only significant correlation between 8.4 GHz A--configuration fluxes
and {\it IRAS} fluxes is found at  25 $\mu$m; there are 
marginally significant correlations at 12 $\mu$m and
60 $\mu$m, but no evidence of a correlation at 100 $\mu$m.
In most cases, the 8.4 GHz A--configuration flux densities of 
Seyferts with a given {\it IRAS} flux density
vary by at least two orders of magnitude. 

In Figure \ref{12-rad.fig} we plot  8.4 GHz A--configuration luminosities 
against {\it IRAS} 12 $\mu$m luminosities.
For a flux--limited sample, correlations between luminosities in
different wavebands can be misleading because of the strong dependence
of luminosity on distance.
In fact, the 8.4 GHz versus 12 $\mu$m luminosity--luminosity plane is
evenly filled between the limits defined by the Malmquist bias and the flux
limits (see Appendix \ref{Malm.sec}) 
and evidence of a true correlation is weak.
Some evidence of a real (steep) dependence of  
8.4 GHz A--configuration luminosity on {\it IRAS} 12 $\mu$m
luminosity is given by the distribution of radio flux density upper--limits,
but this dependence should be judged using partial correlation 
methods \cite{Akritas96} when the effects of sample incompleteness are
properly understood; 
the extended 12 $\mu$m sample was selected using an {\it IRAS}
flux limit of 0.22 Jy, but it is only complete to 0.3 Jy
\cite{RMS93}.

\subsection{Low--resolution radio fluxes}
\label{S1.4-IR.sec}

\begin{figure*} 
\centerline{\includegraphics[angle=0,width=17.5cm]{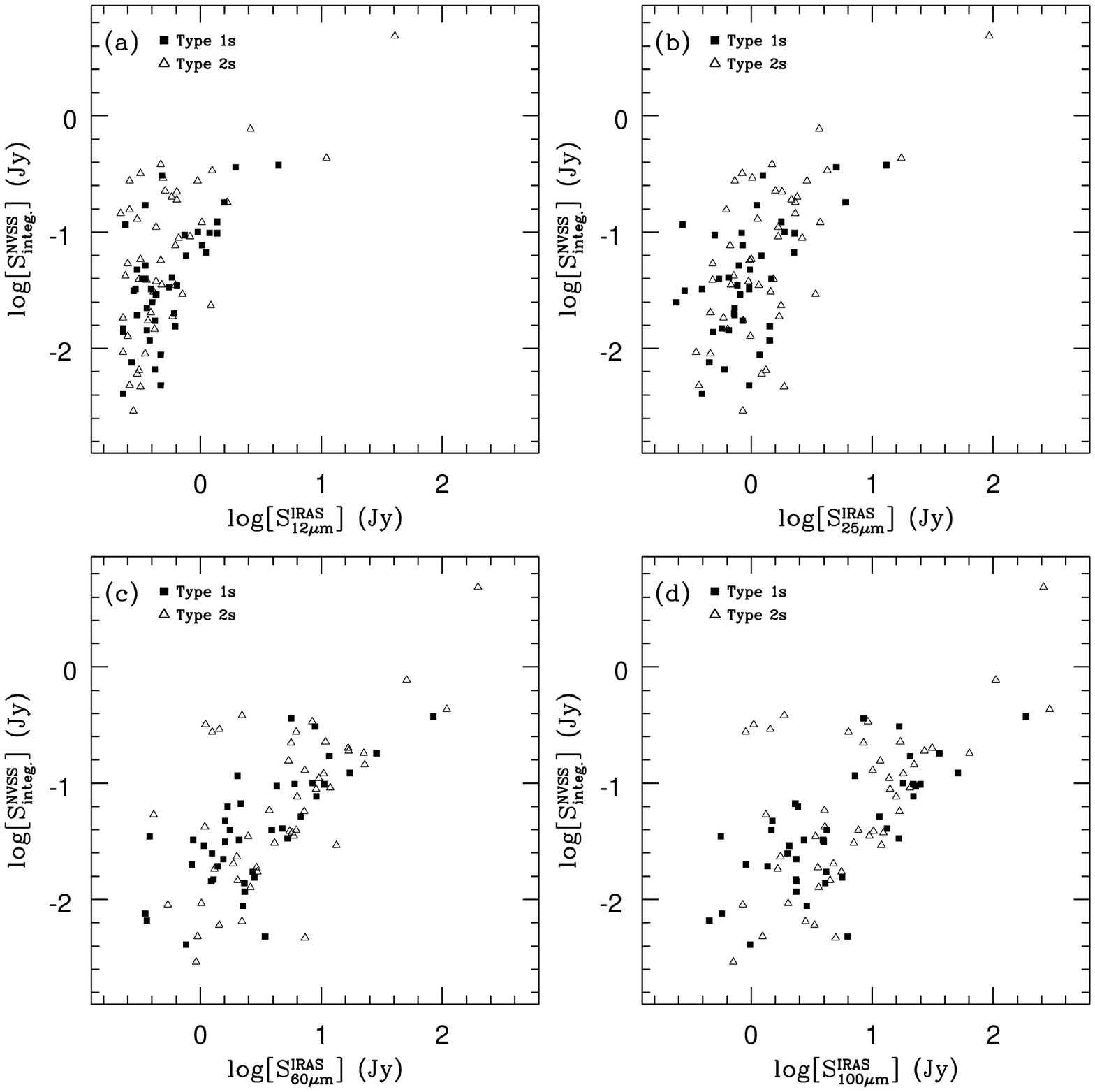}}
\caption{Integrated NVSS radio flux densities are plotted against
(a) 12 $\mu$m, (b) 25 $\mu$m, (c) 60 $\mu$m and (d) 100 $\mu$m
{\it IRAS} flux densities. 
There are extremely significant correlations between NVSS fluxes and
infrared fluxes in each {\it IRAS} waveband.
The group of four type 2 Seyferts centered near
log[S$\rm ^{IRAS}_{100\mu m}$]= 0.1 and log[S$\rm^{NVSS}_{integ.}$]=$-$0.5 in (d) contains three sources with known hidden
broad lines (F01475$-$0740, Markarian 348 and Markarian 463) 
and one which has been classified as both type 1 and type 2 (Markarian 6).
}
\label{IR-SintNvss.fig}
\end{figure*}

\begin{table*}
\scriptsize
\begin{center}
\begin{tabular}{|c|c|r|r|r|r|r|r|} \hline 
{\bf Independent} & {\bf Dependent} & \multicolumn{2}{c|}{\bf Cox} &
\multicolumn{2}{c|}{\bf Kendall's $\tau$} & \multicolumn{2}{c|}{\bf
Spearman's $\rho$} \\ 
{\bf variable} & {\bf variable} & {\bf P(null)} & {\bf $\chi ^{2}$} & {\bf P(null)} & {\bf Z} & {\bf P(null)} & {\bf $\rho$}  \\    
\hline
S$\rm ^{IRAS}_{12\mu m}$   & S$\rm ^{A-con.}_{8.4GHz}$  &33.18\% & ~0.942 & 4.26\% & 2.028 & 5.62\% & 0.194 \\
S$\rm ^{IRAS}_{25\mu m}$   & S$\rm ^{A-con.}_{8.4GHz}$  & 1.47\% & ~5.947 & 0.14\% & 3.203 & 0.21\% & 0.313 \\
S$\rm ^{IRAS}_{60\mu m}$   & S$\rm ^{A-con.}_{8.4GHz}$  &20.85\% & ~1.582 & 2.72\% & 2.209 & 3.97\% & 0.209 \\
S$\rm ^{IRAS}_{100\mu m}$  & S$\rm ^{A-con.}_{8.4GHz}$  &64.90\% & ~0.207 & 9.09\% & 1.691 &13.76\% & 0.151 \\
&&&&&&&\\               
S$\rm ^{IRAS}_{12\mu m}$   & S$\rm ^{NVSS}_{peak}$     & 0.00\% & 16.530 & 0.01\% & 3.922 & 0.02\% & 0.393 \\
S$\rm ^{IRAS}_{25\mu m}$   & S$\rm ^{NVSS}_{peak}$     & 0.00\% & 21.074 & 0.00\% & 4.764 & 0.00\% & 0.470 \\
S$\rm ^{IRAS}_{60\mu m}$   & S$\rm ^{NVSS}_{peak}$     & 0.00\% & 17.404 & 0.00\% & 4.875 & 0.00\% & 0.480 \\
S$\rm ^{IRAS}_{100\mu m}$  & S$\rm ^{NVSS}_{peak}$     & 0.05\% & 12.044 & 0.00\% & 4.250 & 0.01\% & 0.423 \\
&&&&&&&\\               
S$\rm ^{IRAS}_{12\mu m}$   & S$\rm ^{NVSS}_{integ}$   & 0.00\% & 25.471 & 0.00\% & 4.782 & 0.00\% & 0.466 \\
S$\rm ^{IRAS}_{25\mu m}$   & S$\rm ^{NVSS}_{integ}$   & 0.00\% & 23.794 & 0.00\% & 5.044 & 0.00\% & 0.494 \\
S$\rm ^{IRAS}_{60\mu m}$   & S$\rm ^{NVSS}_{integ}$   & 0.00\% & 24.058 & 0.00\% & 5.802 & 0.00\% & 0.553 \\
S$\rm ^{IRAS}_{100\mu m}$  & S$\rm ^{NVSS}_{integ}$   & 0.00\% & 19.370 & 0.00\% & 5.394 & 0.00\% & 0.515 \\
\hline
\end{tabular}
\caption[The probability of correlations between radio fluxes and {\it IRAS} fluxes.]
{High--resolution (A--configuration) and low--resolution (NVSS) radio 
flux densities are compared to infrared flux densities for all 
{\it IRAS} wavebands and the probability that no correlation exists
between radio fluxes and {\it IRAS} fluxes, P(null), is given.
{\it IRAS} fluxes are more closely related to low--resolution 
radio fluxes (NVSS) than high--resolution radio fluxes (A--configuration).
The test statistics ($\chi ^{2}$, Z and $\rho$) show that the 
correlation between {\it IRAS} fluxes 
and NVSS fluxes is stronger for integrated NVSS fluxes than peak NVSS
fluxes in all {\it IRAS} bands (the accuracy of the {\small ASURV} output is
limited to 0.00\%).}  
\label{rad-IR.tab}
\end{center}
\end{table*}

Figure \ref{IR-SintNvss.fig} shows how integrated 
VLA NVSS radio flux densities vary with {\it IRAS} flux densities for the
extended 12 $\mu$m Seyfert sample.  
The probability that NVSS fluxes and {\it IRAS} fluxes are 
correlated is extremely significant for each {\it IRAS} waveband.
Interestingly, correlations between NVSS fluxes and {\it IRAS} fluxes
are stronger for all {\it IRAS} wavebands when integrated
NVSS fluxes are used instead of peak NVSS fluxes.
Table \ref{rad-IR.tab} gives individual test results.

\begin{figure} 
\centerline{\includegraphics[angle=0,width=8.5cm]{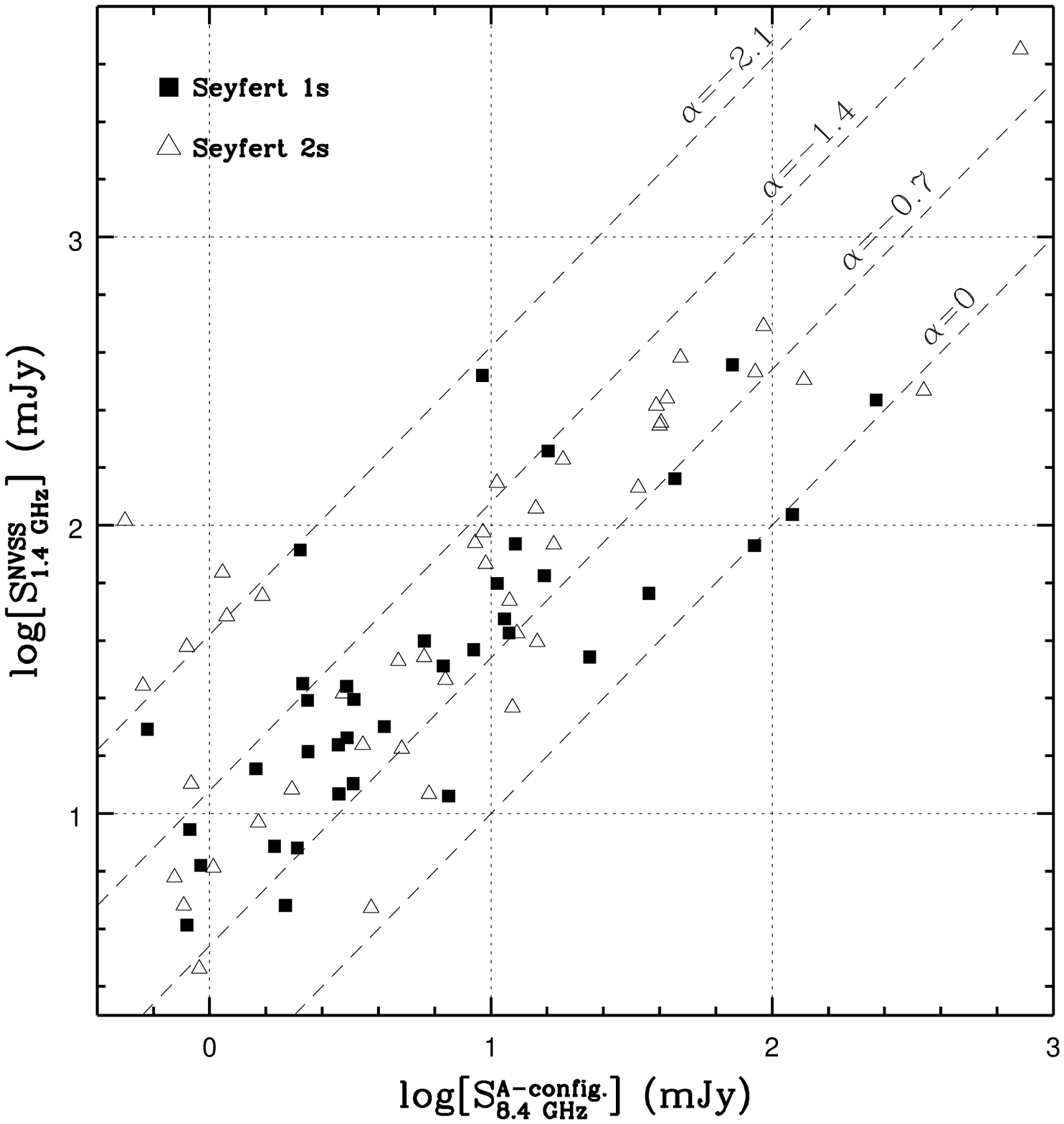}}
\caption{Peak VLA 1.4 GHz NVSS radio flux densities 
are plotted against 8.4 GHz A--configuration flux densities.
Dashed lines indicate the flux relationships expected for various
spectral indices; since NVSS observations measure a higher 
fraction of non--nuclear
emission than 8.4 GHz A--configuration observations, the apparent 
spectral indices measured in this way provide lower limits 
to the true spectral indices of the nuclear components.}
\label{SaSf.fig}
\end{figure}

Figure \ref{SaSf.fig} shows how peak NVSS flux densities vary with 
8.4 GHz A--configuration flux densities.
The two variables are likely to be correlated; the probability of 
the null hypothesis, that there is no correlation 
between the 8.4 GHz fluxes (independent variable) and the 
NVSS peak fluxes, is:
0.50\% (Cox proportional Hazard), 
1.37\% (Kendall's Tau) and 
2.12\% (Spearman's Rho).
For 8.4 GHz fluxes and NVSS integrated fluxes
the probability of the null hypothesis is:
0.56\% (Cox proportional Hazard), 
1.61\% (Kendall's Tau) and 
2.41\% (Spearman's Rho).

The apparent spectral index estimated by comparing  
8.4 GHz A--configuration flux densities and NVSS peak flux densities 
may be regarded as a lower limit on the spectral
index of the active nucleus since NVSS fluxes contain a higher 
fraction of non--nuclear emission than A--configuration fluxes.
The apparent spectral indices measured in this way
lie in the range -1.4 $<\alpha <$ 0 for 80\% of the sources.

\subsection{Implications: {\it IRAS} fluxes are dominated by 
large--scale emission.}
\label{IR_disc.sec}

The correlations between 8.4 GHz A--configuration fluxes and the
fluxes in all {\it IRAS} wavebands (\S \ref{S8.4-IR.sec})
are much weaker than the corresponding relationships between NVSS
fluxes and {\it IRAS} fluxes (\S \ref{S1.4-IR.sec}).
This result indicates that large--scale extra--nuclear emission 
dominates {\it IRAS} fluxes for the Seyferts in the sample.

It is difficult to explain the differences between
the radio to infrared correlations obtained with 1.4 GHz 
NVSS fluxes and 8.4 GHz
A--configuration fluxes solely in terms of differences in frequency,
and the differences between the radio to infrared correlations
obtained with NVSS peak fluxes and NVSS integrated fluxes (\S
\ref{S1.4-IR.sec}) can only be due to 
the effect of spatially distinct flux components;
the fact that correlations between NVSS fluxes and
fluxes at all {\it IRAS} wavelengths are stronger for integrated NVSS
fluxes than for peak NVSS fluxes (Table \ref{rad-IR.tab})
means that a measurable fraction of 
{\it IRAS} emission, including mid--infrared emission, is generated 
by flux components larger than one third of the NVSS resolution i.e.
15 arcsecs  (3.6 kpc at the mean
redshift of the sample).

The most likely reason that NVSS fluxes are more closely related to
{\it IRAS} fluxes than 8.4 GHz A--configuration fluxes is that 
the VLA D and DnC configurations used for the NVSS survey 
are sensitive to low--brightness--temperature radio emission which
is undetectable with the VLA A--configuration in snapshot mode 
at 8.4 GHz i.e. by radio emission with a brightness temperatures 
between around 0.14 K (NVSS limit) and 72 K (8.4 GHz A--configuration 
snapshot limit), 
or from flux components larger than 3.5 arcsec
(1.6 kpc at the mean redshift of the sample).
To some degree, the radio brightness--temperatures 
may be used to distinguish between emission with 
different physical origins e.g.
the VLA in D and DnC configuration is sensitive to radio emission 
from normal H {\small II} regions and supernova remnants in nearby galaxies  
which is not usually compact enough to be detected by 
the VLA in A--configuration at 8.4 GHz \cite{Condon98}.

Our results suggest that large--scale 
extra--nuclear emission regions, rather than
compact nuclear regions, are primarily responsible for the correlation 
between NVSS and {\it IRAS} fluxes; stronger correlations between 8.4
GHz A--configuration fluxes and {\it IRAS} fluxes would be expected if
compact regions dominate. 
We do not favour the interpretation that
high--brightness--temperature radio emission from Seyfert nuclei is driven by 
compact nuclear starbursts \cite{Terlevich92}, because 
such starbursts would be expected to make a significant contribution
to {\it IRAS} fluxes.
Note that for the majority of the sources we have detected,
radio luminosity alone is not a good basis to rule out the
possibility that they are caused by compact starbursts e.g. 
even though the 8.4 GHz A--configuration flux density of the well--known
source ARP 220 would make it the 7th most luminous source in the
sample, \scite{Smith98arp} have provided strong evidence that
its radio emission is dominated by a compact starburst. 

\scite{Baum93} identified three components 
which contribute to the radio emission from Seyfert galaxies;
compact nuclear emission, kiloparsec--scale outflows
and galactic disc emission.
The correlation between radio emission and far-infrared emission is
well known for normal spiral galaxies and starburst galaxies 
(\pcite{deJong85}; \pcite{Helou85}; \pcite{Condon+B86};
\pcite{Condon+B88}; \pcite{Bicay90}) and
is explained by star--formation processes.
Kiloparsec--scale radio outflows extending away from the galactic
plane are found in half or more of Seyfert galaxies and are 
distinguished by their elongated morphology \cite{Colbert96II}.
They are thought to be caused by AGN--driven jets 
but starburst--driven winds cannot be ruled out (\pcite{Baum93};
\pcite{Colbert96II}: \pcite{Colbert98III}).
It is difficult to determine
whether the correlation between NVSS and {\it IRAS} fluxes is  
caused by kiloparsec--scale outflows or galactic disc emission  
because the radio to far--infrared flux ratio of kiloparsec--scale 
outflows is within the range found for normal spiral galaxies and
starburst galaxies \cite{Baum93}.
Since it is an infrared--selected sample, 
the extended 12 $\mu$m sample is likely to contain a high fraction of
galaxies with infrared--bright discs and/or kiloparsec--scale outflows.

Our results are in agreement with those of previous authors who found
that emission from the discs of Seyfert galaxies
makes a significant contribution to their low--resolution radio
fluxes (\pcite{Edelson87}; \pcite{Wilson88}; 
\pcite{RushM+E96};
\pcite{Niklas97}; \pcite{Roy98}) e.g. \scite{RushM+E96} found that 50\%
of Seyfert galaxies have extended flux components larger then 15 arcsec at
5 GHz and estimated that for these sources an average of 33\% 
of the radio emission originates in the galactic disc.
\scite{Rodriguez87} found that active nuclei may not be
responsible for the bulk far-infrared emission from Seyfert galaxies 
at wavelengths longer than 30 microns.
Our results support this interpretation and 
imply that it can be extended to 25 and 12 $\mu$m, at least 
for mid--infrared--selected Seyferts, 
since correlations between radio fluxes and
infrared fluxes at 25 and 12 $\mu$m are much stronger for NVSS fluxes
than for 8.4 GHz A--configuration fluxes. 

If the {\it IRAS} fluxes of Seyfert galaxies 
are dominated by large--scale extra--nuclear regions rather than 
nuclear regions, {\it IRAS} colours do not provide a 
reliable indicator of the inclination of the circum--nuclear torus. 
\scite{Heisler97} made the observation that Seyfert type 2 galaxies 
with hidden broad lines have `hot' {\it IRAS} colours and inferred
that they are sources whose tori have lower inclinations than other
type 2 Seyferts, but an alternative explanation may be necessary 
(see \S \ref{HBLR.sec}).
In addition, if {\it IRAS} colours are not dominated by emission
from dust in the circum--nuclear torus, torus models are of limited
importance in assessing selection biases in {\it IRAS}--selected samples
i.e.  even if the torus is optically--thick at 
mid--infrared wavelengths (\S \ref{bias.sec})
the extended 12 $\mu$m sample may contain representative
populations of type 1 and type 2 Seyferts.

Although {\it IRAS} fluxes seem to be
dominated by emission regions which are less compact that those
responsible for the 8.4 GHz A--configuration emission, 
at least some component of the {\it IRAS} flux at 25
$\mu$m (and possibly at 12 and 60 $\mu$m) is correlated with 
high--brightness--temperature radio emission (\S \ref{S8.4-IR.sec}). 
The reason that 8.4 GHz A--configuration
fluxes are more closely correlated with 25 $\mu$m fluxes 
than with 60 $\mu$m and 100 $\mu$m fluxes may be understood if
high--brightness--temperature radio emission is more closely
related with the hot gas and dust which dominates mid--infrared 
{\it IRAS} emission, than with the cooler gas and 
dust which dominates far--infrared {\it IRAS} emission.
However, this does not explain why 8.4 GHz
A--configuration fluxes are more closely correlated with 25 $\mu$m fluxes 
than 12 $\mu$m fluxes.
The difference may be due to statistical limitations;
Seyferts were selected at 12 $\mu$m, the {\it IRAS} waveband
in which their fluxes are faintest, and 
this means that the {\it IRAS} 12 $\mu$m flux density distribution is the 
most severely truncated and correlations with {\it IRAS} 12 $\mu$m
fluxes are most difficult to quantify.
If the difference is real, it may indicate that 12 $\mu$m emission is
less isotropic than 25 $\mu$m emission; an explanation which is consistent
with models of circum--nuclear tori which are optically--thick
at 12 $\mu$m (see \S\ref{12umFlux.sec}). 
 
The correlation 
between NVSS fluxes and 8.4 GHz A--configuration fluxes can
be explained if, as expected, the compact components 
detected in A--configuration at 8.4 GHz also contribute to the 
NVSS fluxes.
The fact that {\it IRAS} fluxes are 
only weakly related, if at all, to 8.4 GHz A--configuration fluxes
can be explained if 8.4 GHz A--configuration
fluxes are dominated by the active nucleus,
NVSS fluxes contain a mixture of nuclear emission and extra--nuclear
emission, 
and {\it IRAS} fluxes are dominated by extra--nuclear emission.
This explanation is favoured by the fact that the correlation 
between NVSS fluxes and 8.4 GHz A--configuration fluxes is stronger
for peak NVSS fluxes than integrated NVSS fluxes.

\section{HOST GALAXIES}
\label{host.sec}

\subsection{Large--scale emission from type 1 and type 2 Seyferts}
\label{S1.4-IR12.sec}

\begin{figure} 
\centerline{\includegraphics[angle=0,width=8.5cm]{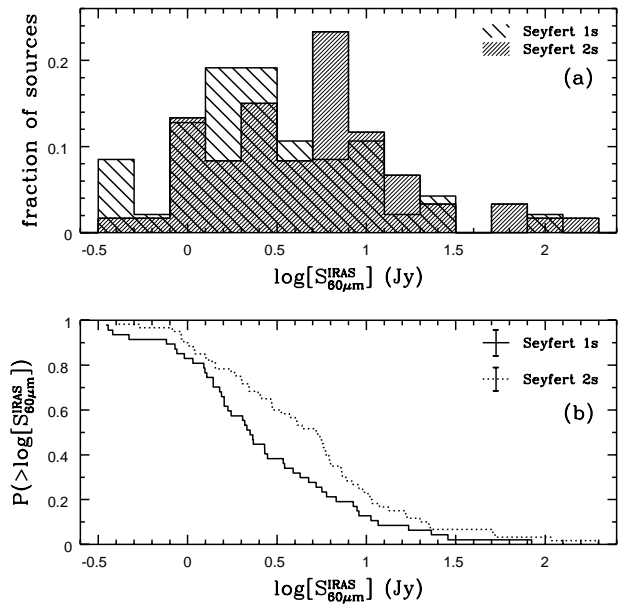}}
\caption{(a) Histograms 
showing the fractional {\it IRAS} 60 $\mu$m flux density
distributions and (b) the cumulative  {\it IRAS} 60 $\mu$m flux density
distributions of 47 type 1 and 60 type 2 Seyferts in the extended
12 $\mu$m sample. 
Type 2 Seyferts have significantly higher 
{\it IRAS} 60 $\mu$m fluxes than type 1 Seyferts.}
\label{S60.fig}
\end{figure}

\begin{figure} 
\centerline{\includegraphics[angle=0,width=8.5cm]{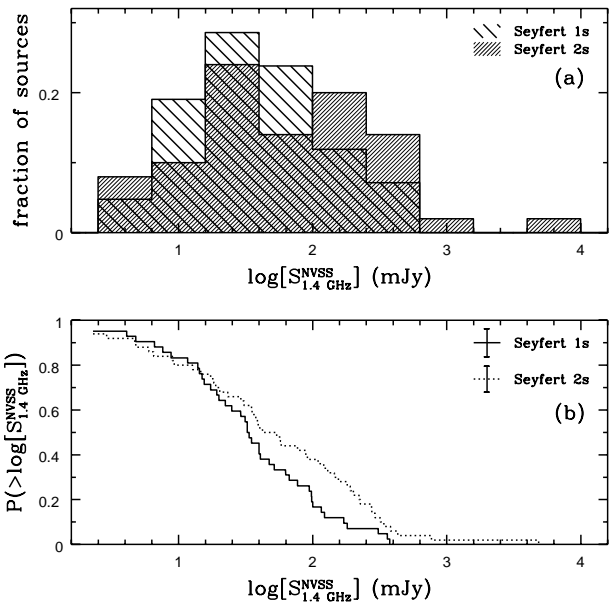}}
\caption{(a) Histograms 
showing the fractional NVSS integrated flux density
distributions and (b) the cumulative NVSS integrated 
flux density distributions of 40 type 1 and 47 type 2 Seyferts in the extended
12 $\mu$m sample.
The NVSS integrated flux density distributions of the two
Seyfert types are indistinguishable.}
\label{NVSSflux.fig}
\end{figure}

Figure \ref{S60.fig} shows the 
{\it IRAS} 60 $\mu$m flux density distributions of 
type 1 and type 2 Seyferts in the extended 12 $\mu$m sample.
Type 2 Seyferts are significantly brighter than type 1 Seyferts.
The probability of the null hypothesis, that the 60 $\mu$m fluxes 
of the two Seyfert types are drawn at random from the same 
parent population, is: 4.66\% (KS), 
2.16\% (Gehan's, perm.), 1.93\% (Gehan's, hyper.), 4.10\% (Logrank test)
and 4.10\% (Peto \& Peto).  

Figure \ref{NVSSflux.fig} shows the NVSS integrated flux density 
distributions of type 1 and type 2 Seyferts in the sample.
Type 2 Seyferts tend to have brighter NVSS fluxes, but the two 
flux distributions are statistically indistinguishable.
The probability of the null hypothesis, 
that the NVSS integrated flux density distributions of the two Seyfert types 
are drawn at random from the same parent population, is: 19.45\% 
(Gehan's, perm.), 19.68\% (Gehan's, hyper.), 64.63\% (Logrank test),
19.45\% (Peto \& Peto) and 19.65\% (Peto \& Prentice).  

As a crude estimate of the strength of extended radio components 
we have calculated the quantity S$_{\rm ext}$ by 
subtracting 8.4 GHz A--configuration flux densities from integrated NVSS 
flux densities.
Type 2 Seyferts tend to have higher S$_{\rm ext}$ values than type 1
Seyferts, but the difference between the S$_{\rm ext}$ distributions 
is only marginally significant.
The probability of the null hypothesis, that S$_{\rm ext}$ 
values of type 1 and type 2 Seyferts are drawn at random from 
the same parent population, is: 
11.31\% (Gehan's, perm.), 11.04\% (Gehan's, hyper.), 
2.87\% (Logrank test) and 2.87\% (Peto \& Peto).

\subsection{Implications: more extended emission in type 2 Seyferts ?}
\label{star2.sec}

Our results, which suggest that type 2 Seyferts have brighter 
{\it IRAS} 60 $\mu$m fluxes than type 1 Seyferts, are not expected 
according to the Seyfert unification model.

One possibility is that the extended 12 $\mu$m Seyfert sample is biased
against low--luminosity type 2 Seyferts; for example if {\it IRAS} 
emission is dominated by circum--nuclear tori which are 
optically--thick at 12 $\mu$m, or if the population of
newly--recognised type 2 Seyferts identified in Section \ref{ID.sec} 
are low--luminosity sources. 
However, we do not favour this explanation because
{\it IRAS} fluxes appear to be dominated by extra--nuclear emission
regions (\S \ref{IR.sec}) and
the sample appears to contain well--matched 
populations of type 1 and type 2 Seyferts (\S \ref{bias.sec}).

If extra--nuclear emission regions dominate {\it IRAS} fluxes, 
our results imply that they are brighter or more common in
type 2 Seyferts than type 1 Seyferts.
The marginally significant trend for type 2 Seyferts to produce more
large--scale radio emission (S$_{\rm ext}$) than type 1 Seyferts 
favours this explanation.
Note that differences between the integrated NVSS fluxes of type 1 and type 2
Seyferts will be hard to distinguish if  
NVSS fluxes contain a high fraction of AGN--related emission 
(\S \ref{IR_disc.sec}).
An excess in the extra--nuclear {\it IRAS} and NVSS emission 
of type 2 Seyferts supports claims that the host galaxies of type 2
Seyferts show excess of star--formation
(\pcite{Maiolino95a_Ia}; \pcite{Hunt99I}) and an excess of wisps,
filaments and dust lanes \cite{Malkan98}.
Another alternative is that some 
starburst galaxies in the extended 12 $\mu$m
sample are mis--classified as type 2 Seyferts.

\subsection{Host galaxy morphological types}
\label{T.sec}

\begin{figure} 
\centerline{
       \includegraphics[angle=0,width=8.5cm]{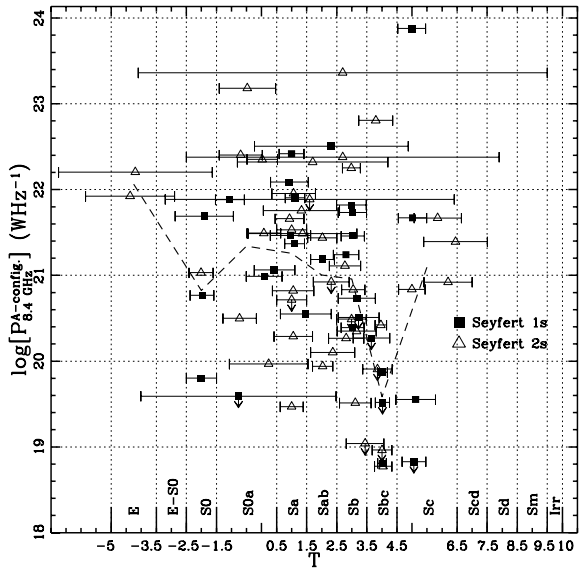}}
\caption{The 8.4 GHz VLA A--configuration radio powers of Seyferts from
the extended 12 $\mu$m sample are likely to be anti--correlated with the 
morphological type of the host galaxy, T, 
especially in the interval $-$1.5$<$T$<$4.5. 
A link between galaxy evolution and nuclear activity is indicated.
Hubble types which correspond to the various ranges in
T are shown on the x axis. 
The dashed line joins the mean values for each Hubble type 
(where the minimum value in a bin is an upper--limit, it is treated as
a detection and the estimate of the mean is biased towards high values).   
The four most luminous sources, in order of decreasing luminosity, are:
Markarian 231, Markarian 463, Markarian 348 and Markarian 533. 
}
\label{t-P.fig}
\end{figure}

\begin{figure} 
\centerline{
       \includegraphics[angle=0,width=8.5cm]{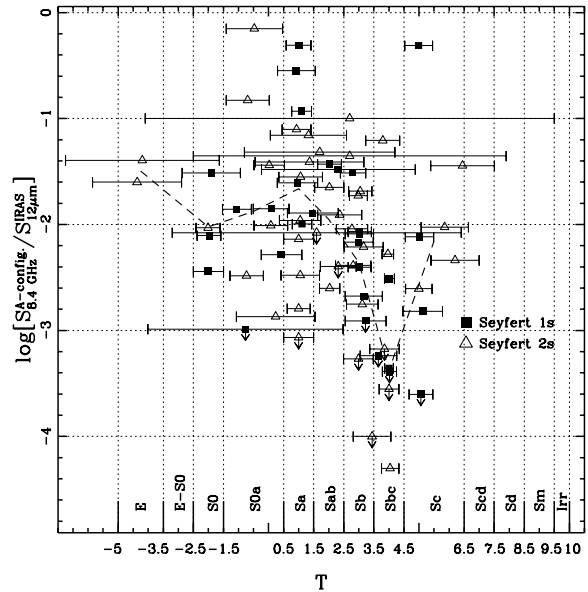}}
\caption{The `radio--loudness' of Seyferts from
the extended 12 $\mu$m sample, as indicated by the 8.4 GHz VLA
A--configuration to IRAS 12 $\mu$m flux ratio, 
is likely to be anti--correlated with the morphological
type of the host galaxy, T, 
especially in the interval $-$1.5$<$T$<$4.5.
The dashed line joins the mean values for each Hubble type. 
}
\label{t-rad12.fig}
\end{figure}

Figure \ref{t-P.fig} shows how the 8.4 GHz A--configuration
luminosities of the extended 12 $\mu$m sample Seyferts are 
distributed according to the host morphological type, T.
If radio luminosity upper--limits are considered, 
it is likely that there is an anti--correlation between T and radio power.
The probability of the null hypothesis, that there is no correlation
between host morphology (independent variable) and 8.4 GHz
A--configuration luminosity, is:
0.78\% (Cox proportional Hazard), 0.83\%
(Kendall's Tau) and 0.87\% (Spearman's Rho).
The anti--correlation is only statistically
significant because of a group of late--type galaxies which were 
undetected in our radio survey, 
without the inclusion of luminosity upper--limits the 
probability of the null hypothesis is: 65.56\% (Cox proportional
Hazard), 8.59\% (Kendall's Tau) and 9.29\% (Spearman's Rho).
On the other hand, the anti--correlation is highly significant for 
the 58 sources 
with morphological types in the range $-$1.5$<$T$<$4.5 
(Hubble types S0a to Sbc), for this sub--population  
the probability of the null hypothesis is:
0.15\% (Cox proportional Hazard), 0.08\%
(Kendall's Tau) and 0.12\% (Spearman's Rho).
Note that the statistical tests used do not take into account  
uncertainties in morphological type.

Figure \ref{t-rad12.fig} shows how a measure of `radio--loudness', 
the 8.4 GHz to {\it IRAS} 12 $\mu$m flux ratio, is 
distributed according to the host morphological type, T.
Even without the inclusion of flux density upper--limits, there is a
marginally significant probability that there is an anti--correlation
between T and the 8.4 GHz to {\it IRAS} 12 $\mu$m flux ratio.
The probability of the null hypothesis (no correlation) when 
radio flux density upper--limits are included is:
0.49\% (Cox proportional Hazard), 0.15\% (Kendall's Tau) and 
0.22\% (Spearman's Rho).
Without the inclusion of radio flux density upper--limits the probability 
of the null hypothesis is: 14.15\% (Cox proportional Hazard), 1.52\%
(Kendall's Tau) and 2.19\% (Spearman's Rho).
For sources in the range $-$1.5$<$T$<$4.5 the probability 
of the null hypothesis is:
0.46\% (Cox proportional Hazard), 0.04\%
(Kendall's Tau) and 0.11\% (Spearman's Rho).

There is a statistically significant trend for galaxies with uncertain 
morphological types to host more powerful radio sources than other galaxies.
There are 12 sources for which the uncertainty on the 
morphological type code is greater than unity, 
and the probability that the radio powers of these sources are drawn
at random from the sample is low;
2.41\% (Gehan's, perm.), 0.78\% (Gehan's, hyper.), 3.46\%
(Logrank), 2.65\% (Peto \& Peto) and 1.48\% (Peto \& Prentice).
The probability that the 8.4 GHz to {\it IRAS} 12 $\mu$m flux ratio
of these sources are drawn from the sample at random is: 
3.68\% (Gehan's, perm.), 1.56\% (Gehan's, hyper.),  8.06\%
(Logrank), 3.86\% (Peto \& Peto) and 2.24\% (Peto \& Prentice).
This trend may reflect observational difficulties 
in determining the morphological types of distant galaxies rather
than a true link between morphological ambiguities and nuclear
activity i.e. due to the Malmquist bias,  
distant sources are likely to be more powerful than nearby ones.
Galaxies for which the uncertainty on the morphological type code 
is greater than unity tend to be more distant than other 
sources in the sample, but this trend is of marginal significance. 
The probability that galaxies for which the uncertainty on the 
morphological type code is greater than unity have the same redshift
distribution as the remaining sources is: 
7.90\% (Gehan's, perm.), 9.77\% (Gehan's, hyper.), 3.09\%
(Logrank) and 3.09\% (Peto \& Peto).

\subsection{Implications: the strength of a nuclear radio source is
related to host galaxy morphology.}

There is a large degree of scatter in the  
formally significant anti--correlation between nuclear radio luminosity 
and host galaxy morphological type for the Seyferts in the extended 12
$\mu$m sample, and its significance relies on luminosity upper--limits. 
A possible relationship between nuclear radio power and host galaxy
morphology is highlighted by the sub--population of  Sbc galaxies in 
the sample, which tend to have weaker nuclear radio sources than other
types of galaxy.
Evidence of an anti--correlation between the strength of nuclear activity 
and host galaxy morphology is clearest for sources with intermediate
Hubble types (S0a to Sbc) and when a measure of
radio--loudness, the radio to 12 $\mu$m {\it IRAS} flux ratio,
is considered instead of radio luminosity alone. 
The reason galaxies with morphological types 
earlier than S0a weaken the significance of the anti--correlation
between radio power and host galaxy morphology may be related to the
effects of mergers; the two galaxies 
classified as ellipticals (NGC 1143/4 and NGC 4922A/B) 
are multiple systems which are probably in the process of merging.
The reason that galaxies with morphological types 
later than Sbc weaken the significance of the anti--correlation  
between radio power and galaxy morphology is unclear.

It is possible that our results are affected by bias; 
the selection criteria used to define the extended 12
$\mu$m sample could mean that galaxies with low--luminosity AGN may
satisfy the selection criteria simply because they have
mid--infrared--bright discs, and this may introduce a bias
against low--luminosity AGN in early--type galaxies.
On the other hand, 
the trend for early--type galaxies to host 
more powerful radio sources
than late--type galaxies is consistent with the fact that radio--loud
AGN are usually found in elliptical galaxies and the fact
that Seyfert nuclei are found preferentially in early--type host
galaxies (\pcite{Moles}; \pcite{Hunt99I}).
Previous studies have shown that galaxies with luminous bulges have the most
powerful nuclei for both active galaxies
(\pcite{Whittle92III}; \pcite{Ho99}) and
normal galaxies (\pcite{Sadler89}; \pcite{Kormendy+R95}).
Taken together, recent results which link the mass of the central
supermassive black hole of a galaxy to its bulge mass
\cite{Magorrian98} and the radio power of its nucleus 
\cite{Franceschini98} suggest that a correlation between bulge mass
and nuclear radio power is expected, and it is possible that 
the loose relationship between 
nuclear radio power and galaxy morphological type that we have found
is indicative of a more fundamental relationship of this type.

In Section \ref{IR.sec} we have shown that nuclear radio fluxes are more
closely related to {\it IRAS} fluxes at 25 $\mu$m than at any other 
{\it IRAS} wavelength and therefore 
our results are in good agreement with the anti--correlation between 
25 $\mu$m to 60 $\mu$m {\it IRAS} flux ratios and morphological type
found for extended 12 $\mu$m Seyferts \cite{Hunt99I}.

If there is an anti--correlation between radio power and morphological type,
T, one source which does not fit the general trend is the
ultra--luminous infrared galaxy Markarian 231. 
This galaxy, which has an Sc morphology and a powerful, 
compact nuclear radio source, is probably the most luminous 
AGN in the local universe \cite{Sanders88}.
In Section \ref{compactImp.sec} we considered the possibility 
that it contains a 
young radio source which may be in a particularly luminous phase 
of evolution.

\subsection{Host galaxy orientation}
\label{PA.sec}

\begin{figure} 
\centerline{
       \includegraphics[angle=0,width=8.5cm]{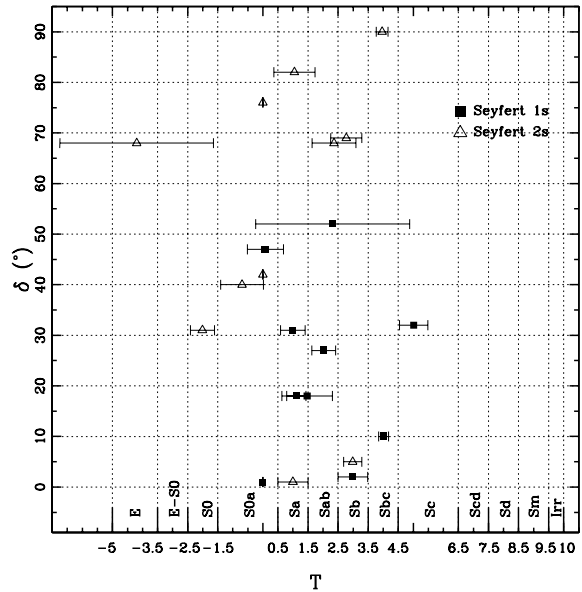}
}
\caption{The difference between the position angles of the host galaxy
major axes and the position angles of collimated nuclear radio
structures, $\delta$, is plotted against the morphological
type of the host galaxy, T.
There is no apparent relationship between these two quantities.}
\label{delPA.fig}
\end{figure}

The difference between the position angle of the host galaxy
major axis and the position angle of a collimated nuclear radio
structure, $\delta$, is an indicator of how well the central AGN is
aligned with the rotation axis of the host galaxy;
$\delta$ is 90 degrees
for a central engine which is aligned with the galaxy rotation axis
(the reliability with which $\delta$ indicates the relative
orientation of the AGN and the galactic plane decreases 
with galaxy inclination, but all galaxies for which values of $\delta$ 
are available have an inclination greater than 30 degrees). 

Figure \ref{delPA.fig} shows a plot of 
$\delta$ versus host galaxy morphological type, T.
We find an even distribution of $\delta$ for the sample as a
whole, suggesting that Seyfert central engines are randomly oriented
with respect to their host galaxies.
There is a marginally significant trend for type 2 Seyferts to 
show higher values of $\delta$ than type 1 Seyferts, but this difference 
relies on a small number of sources.
The probability of the null hypothesis, that the $\delta$
distributions of type 1 and type 2 Seyferts are drawn at random 
from the same parent population, is:
14.20\% (Gehan's, perm.), 13.83\% (Gehan's, hyper.), 2.42\%
(Logrank) and 2.42\% (Peto \& Peto).
There is no obvious relationship between $\delta$ and T. 
The probability that there is no correlation between $\delta$ and T is:
90.66\% (Cox proportional Hazard), 83.20\%
(Kendall's Tau) and 71.16\% (Spearman's Rho).

\subsection{Implications: no alignment between the central engine and 
the galactic disc.}

Our results suggest that Seyfert central engines are randomly oriented
with respect to their host galaxies.
This result is in agreement with studies of individual sources
whose 3--dimensional geometry can be measured accurately 
(e.g. \pcite{Miyoshi95})
and other statistical studies using radio observations 
(\pcite{Schmitt97}; \pcite{Nagar99PA}).
The main advantage of our study is that our sources are drawn from a 
homogeneously--selected sample.

Previous authors 
have found weak evidence that there is a deficiency of 
type 2 Seyferts with high values of $\delta$, the
difference between the position angle of the host galaxy
major axis and the position angle of a collimated nuclear radio
structure (\pcite{Schmitt97}; \pcite{Nagar99PA}).
In contrast to these results, we find a marginally significant 
deficiency of type 1 Seyferts with high values of $\delta$. 
Our result, together with those of previous authors, probably implies that
there is no real difference between the $\delta$ distributions of type
1 and type 2 Seyferts.
The low--significance differences between 
the $\delta$ distributions of type 1 and type 2 Seyferts which have
been found to date may be the result of small--number statistics.

We find no evidence to support the weak trend for 
late--type Seyferts galaxies to favor larger values of $\delta$ 
found by \scite{Wilson+Tsvet94} and \scite{Nagar99PA}.

\section{NUCLEAR OUTFLOWS AND V--SHAPED EXTENDED EMISSION--LINE REGIONS}
\label{Cone.sec}

\begin{figure} 
\centerline{
       \includegraphics[angle=0,width=8.5cm]{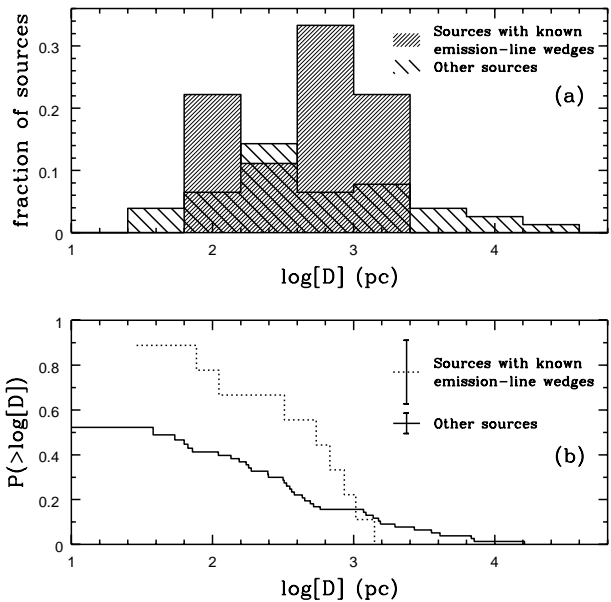}
}
\caption{(a) Histograms showing the fractional radio size
distributions of the 8 
extended 12 $\mu$m Seyferts which are known to contain V--shaped 
extended emission--line regions and were resolved 
and the 36 other resolved sources in the sample. 
(b) The cumulative size distributions of the 9 sources 
with V--shaped extended emission--line regions which were 
detected and the 77 other detected sources,
as given by the Kaplan--Meier estimator where the y axis gives the
probability that a source is larger than a given size.
Sources known to have V--shaped extended emission--line regions are
significantly larger than other sources.}
\label{con.fig}
\end{figure}

\scite{Mulchaey96II} listed 15 Seyferts whose
Extended Narrow--Line Regions (ENLRs) have a V--shaped morphology
consistent with a cone--shaped nuclear radiation field.
The following nine sources are common to the extended 12 $\mu$m
sample (the type of radio structure observed 
in A--configuration at 8.4 GHz is given in parenthesis): 
NGC 262=Markarian 348 (U), NGC 526A (S), NGC 1068 (L), NGC 1365 (A), 
Markarian 6 (L), 
NGC 4151 (L), NGC 4258=Markarian 766 (S), NGC 4388 (L) and NGC 7582 (D).
Since emission--line images of all sources are
not available it is unlikely that these are the only sources 
in the sample with V--shaped ENLRs. 

Only one source with a V--shaped ENLR, Markarian 348, is
unresolved at our resolution and 
this source is known to contain a linearly--aligned triple radio structure
at higher resolution (\pcite{Neff83}; \pcite{Unger84}). 
Another source, which is only slightly resolved 
at 0.24 arcsec resolution, NGC 4528 (Markarian 766), is found to contain
a slightly curving triple radio structure at higher resolution
(unpublished MERLIN 5 GHz observations).

Sources known to contain V--shaped ENLRs have significantly 
larger radio sources than other types of source.
Figure \ref{con.fig} shows the size distributions 
of the 9 extended 12 $\mu$m Seyferts with V--shaped ENLRs compared to the 
remaining sources in the sample.
The probability of the null hypothesis, that the size distributions of 
sources with known V--shaped ENLRs and other sources are drawn
at random from the same parent population, is:
3.00\% (Gehan's, perm.), 1.80\% (Gehan's, hyper.), 0.73\% (Logrank), 2.78\%
(Peto \& Peto) and 3.78\% (Peto \& Prentice).
For a bright subpopulation of 46 sources with 8.4 GHz A--configuration flux 
densities greater than 5 mJy/beam, the probability of the null hypothesis is:
21.46\% (Gehan's, perm.), 19.52\% (Gehan's, hyper.), 16.57\% (Logrank), 22.04\% (Peto \& Peto) and 23.54\% (Peto \& Prentice).

Sources with V--shaped ENLRs are drawn from a nearby
sub--population of the sample.
The probability of the null hypothesis, that the redshift 
distributions of sources with known V--shaped ENLRs and 
other sources are drawn at random from the same parent population, is: 
3.48\% (Gehan's, perm.), 1.39\% (Gehan's, hyper.), 0.44\% (Logrank)
and 0.44\% (Peto \& Peto).
In fact, none of the sources with a known V--shaped ENLR
is at a redshift higher than 0.020, and this probably 
reflects the scope of emission--line imaging surveys carried out to
date and the difficulty of detecting faint extended emission--line regions
in distant sources.
For the sample as a whole, linear radio sizes tend to increase 
with redshift and therefore, if the fraction of nearby sources with both
collimated radio structures and V--shaped ENLRs is similar for distant
sources,
the true significance of the difference in size between sources with
V--shaped ENLRs and other Seyferts is likely to be greater than 
indicated by our comparison.
Of a volume--limited sub--sample of sources closer than redshift
0.020, V--shaped ENLRs have been found in 
four from nine sources with type L radio structures and two from nine
sources with type S radio structures.

The connection between radio outflows and emission--line
wedges is strengthened by radio observations of the six 
sources from the list of \scite{Mulchaey96II} 
which are not found in the extended 12 $\mu$m sample.
Five of these sources 
have linear radio structures when observed at similar or identical
resolution to our observations: Markarian 573 and NGC 5252
\cite{Kukula95}, Circinus \cite{Elmouttie98}, NGC 5728 
\cite{Schommer88} and Markarian 78 \cite{Pedlar.mrk78}.
Only one (NGC 3281) has an unresolved radio structure
\cite{Ulvestad+W89}, 
and this means that elongated radio outflows indicative of radio jets 
are known for all but 3 sources  
from the list of \scite{Mulchaey96II}: NGC 1365, NGC 3281 and NGC 7582.

\subsection{Implications: extended emission--line wedges
are associated with radio outflows.}

We have found that sources with emission--line wedges have 
significantly larger nuclear radio structures than other Seyferts. 
Given that extended emission--line wedges only appear to be present in around
one in every eight Seyfert galaxies \cite{Mulchaey96II} 
and compact, collimated radio structures are found in around one in every five
Seyfert galaxies (\S \ref{D.sec}), 
the fact that these observational features often coincide, making comparisons between 
their orientations possible in the majority of cases (\pcite{Wilson+Tsvet94}; \pcite{Nagar99}), 
is probably significant.

The main limitation of our comparison is
that emission--line images are not available for  
all sources in the sample, and since sources with elongated 
radio structures were specifically
targeted in early ENLR searches (radio maps were relied on to determine 
slit orientations using slit spectroscopy), we cannot rule out
a bias against the identification of V--shaped
ENLRs in sources with unresolved radio sources.
The rate of occurrence of clear V--shaped ENLRs in the 
well--identified Early--type sample \cite{Mulchaey96II}
implies that around 5 V--shaped ENLRs remain to be 
identified in the extended 12 $\mu$m Seyfert sample.
Since the sub--population of sources with V--shaped ENLRs
is small, the addition of 5 sources with unresolved radio structures
would be enough to make the radio size distribution of sources with
V--shaped ENLRs match that of other Seyferts.
On the other hand, sources with V--shaped ENLRs are more nearby than
other types of Seyfert and this suggests that incompleteness is
caused by difficulties in identifying V--shaped ENLRs in distant
sources, rather than sources with unresolved radio structures.
In fact, since distant sources tend to have larger linear sizes than nearby
sources, a complete emission--line survey of the full sample
may increase the significance of our result.

A relationship between extended radio sources and extended 
emission--line regions does not follow directly from the standard 
Seyfert unification model.
According to the unification model a V--shaped ENLR is the result of 
a dusty circum--nuclear torus shadowing a central, isotropic uv source and
the size and morphology 
of the central radio source is not an important parameter.
One proposition which could reconcile our results with the unification
model is that large--scale collimated outflows are found
preferentially in sources with luminous uv sources, in which 
ENLRs are most easily recognised.
Correlations between radio power and [O{\small III}] luminosity are known for 
Seyfert galaxies (\pcite{deBruyn+W78};
\pcite{Whittle92III}) and 
resolved radio sources 
are, on average, more luminous than unresolved sources (\S \ref{D.sec}).
The fact that the radio structures of 
sources with V--shaped emission line regions are not
significantly larger than other sources in a radio--bright sub--sample
of sources supports this interpretation, but a true size difference would also
be less significant in this smaller sample.
An analysis of the emission--line fluxes of the sample, and further
emission--line imaging, would help to determine whether our result can
be explained by selection effects.

If there is a true connection between large
radio sources and V--shaped extended emission--line regions, one
possibility is that it is caused by interactions between nuclear 
outflows and the nuclear environment.
Nuclear outflows of the size we have observed are unlikely to play
a direct r\^{o}le in compressing/ionising extended emission--line 
regions because the ionised gas usually extends twice to five times as far 
from the active nucleus as the nuclear radio jets \cite{Nagar99}.
However, kiloparsec--scale radio outflows which are unobservable at
high resolution are found in as many as half of all Seyferts 
\cite{Colbert96II} and are thought to be caused by AGN--driven
jets (\pcite{Baum93}; \pcite{Colbert96II}; \pcite{Colbert98III}).
The r\^{o}le of such outflows in evacuating and compressing ENLR gas
would help explain the hollow ionisation
cones found in certain sources (\pcite{Pedlar.mrk78}; 
\pcite{Wilson93}; \pcite{Christopoulou1997}).

Another possibility is that nuclear outflows 
ease the passage of ionising nuclear photons to extra--nuclear regions.
\scite{Schulz88} suggested that extended narrow--line regions 
could be explained by holes in the narrow--line region caused by 
radio jets which push aside, or destroy, obscuring material.
Recent HST observations support the idea that nuclear
outflows evacuate channels in the narrow--line region and ease the
passage of nuclear photons to extra--nuclear regions 
(\pcite{Falcke98}; \pcite{Axon98}; \pcite{Capetti99}). 
\scite{Wilson+Tsvet94} pointed out that radio ejecta are more 
tightly collimated than extended emission--line wedges and reasoned
that long, narrow channels in the narrow--line
region are not responsible for collimating the photon field which ionises
the ENLR. 
However, the channels evacuated by 
narrow radio outflows are considerably wider than the radio jets themselves
\cite{Capetti99}, and even a relatively narrow outflow can 
cause a wide--angled uv--cone if 
the most optically thick region is close to the nucleus e.g. if
circum--nuclear tori are created by nuclear outflows.

Lastly, a connection between radio outflows and
extended emission--line regions could be related to source evolution. 
If Seyferts with large radio outflows are older 
than other types of Seyfert (\S \ref{compactImp.sec}) our results imply that 
extended emission--line wedges are also more likely to be found in old
sources. 
Note that nuclear photons which travel to the ends of 
kiloparsec--scale emission--line regions are around 1$\times$10$^{4}$
years old and in Section \ref{compactImp.sec} 
we discussed evidence that the largest (perhaps oldest)
radio outflows from Seyferts are only around an order of magnitude
older.

\section{SEYFERTS WITH HIDDEN BROAD--LINE REGIONS}
\label{HBLR.sec}

\begin{figure} 
\centerline{
       \includegraphics[angle=0,width=8.5cm]{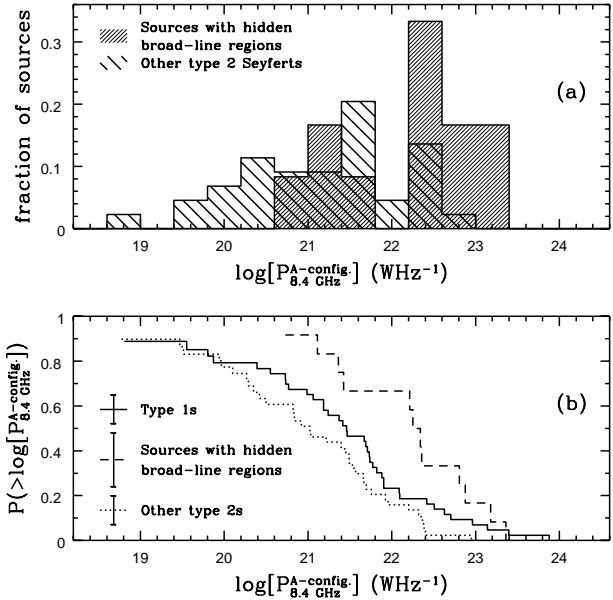}
}
\caption{
(a) The fractional 8.4 GHz A--configuration luminosity 
distributions of the 12 sources in the 
extended 12 $\mu$m sample with known hidden broad line regions
and the other 37 type 2 sources detected.  
(b) The cumulative 8.4 GHz A--configuration 
luminosity distributions of all observed sources 
(42 type 1s, 12 known hidden--broad--line sources and 44 other type 2s)
as given by the Kaplan--Meier estimator, where the y axis gives the  
probability that a source is more luminous than a given radio power. 
Sources known to contain hidden broad line regions 
have more powerful nuclear radio sources than both type
1 sources and other type 2 sources.}
\label{lPaHBLR.fig}
\end{figure}

\begin{figure} 
\centerline{
       \includegraphics[angle=0,width=8.5cm]{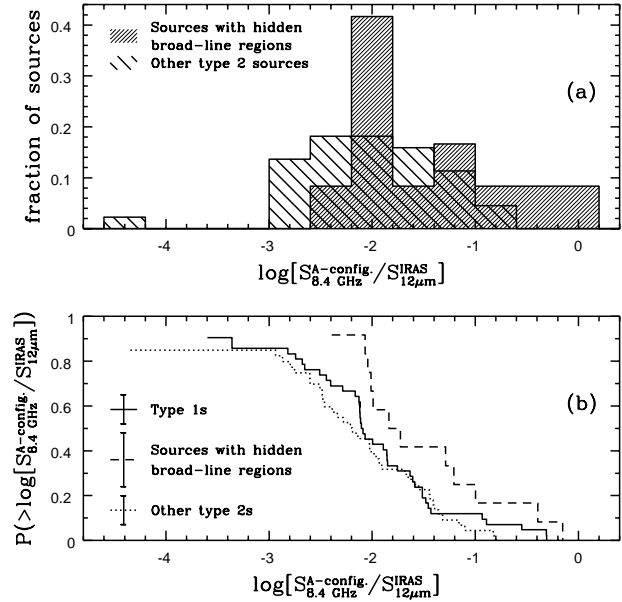}
}
\caption{(a) The fractional distributions of the 
8.4 GHz to {\it IRAS} flux ratio  
for the 12 sources in the extended 12 $\mu$m sample
with known hidden broad line regions 
and other 37 type 2 sources detected. 
(b) The cumulative distributions of the 8.4 GHz to {\it IRAS} flux
ratios of all observed sources 
(42 type 1s, 12 known hidden--broad--line sources and 44 other type 2s)
as given by the Kaplan--Meier estimator.
Sources which are known to show 
hidden broad lines have a higher 8.4 GHz VLA A--configuration to
{\it IRAS} 12 $\mu$m flux ratios than both type
1 sources and other type 2 sources.   
}
\label{rHBLR.fig}
\end{figure}

\scite{Dopita98} identified the following 
14 sources in the extended 12 $\mu$m sample which were classified as
type 2 Seyferts by \scite{RMS93}, and which are known to contain
hidden broad lines (the type of radio structure observed 
in A--configuration at 8.4 GHz is given in parenthesis): 
F00198-7926 (unobserved), NGC 262=Markarian 348 (U), F01475-0475 (U),
NGC 1068 (L), F04385-0828 (U), F05189-3524 (S),
NGC 4388 (L), TOL 1238-364 (A), MCG-3-34-63 (S), Markarian 463 (L),
F15480-0344 (U), IC 5063 (unobserved), F22017+0319 (L) and NGC
7674=Markarian 533 (L). 
Since this list has been drawn from the
literature it is unlikely to be complete.

We find that sources with known hidden broad--line regions contain 
radio sources which are significantly more powerful than other
Seyferts in the sample. 
A high fraction of the most radio luminous sources in the sample are
either type 1 Seyferts or type 2 Seyferts with known hidden broad lines: 
of the 20 most radio luminous 
sources in the sample only 5 are not known to contain broad--line regions
(Markarian 273, Markarian 6, F08572+3915, NGC 1125 and NGC 5256), and 
of the 20 sources with the highest 8.4 GHz to {\it IRAS} 12 $\mu$m 
flux ratios only 2 are not known to contain broad--line regions 
(Markarian 6 and NGC 5506).

Figure \ref{lPaHBLR.fig} shows the 8.4 GHz A--configuration luminosity
distributions of sources with known hidden broad--line regions and other
type 1 and type 2 Seyferts from the extended 12 $\mu$m sample.  
Sources with known hidden broad--line regions are significantly more 
luminous than both type 1 Seyferts and type 2 Seyferts not known to 
show hidden broad--line regions.
The probability of the null hypothesis, that the luminosities
of type 1 sources and sources with known hidden broad--line regions  
are drawn at random from the same parent population, is: 
3.02\% (Gehan's, perm.), 1.61\% (Gehan's, hyper.), 1.19\% 
(Logrank test), 3.03\% (Peto \& Peto) and 2.85\% (Peto \& Prentice).  
The probability of the null hypothesis, that the luminosties of
type 2 Seyferts not known to show hidden broad--line regions 
and type 2 Seyferts with known hidden broad--line regions  
are drawn at random from the same parent population, is: 
0.15\% (Gehan's, perm.), 0.02\% (Gehan's, hyper.), 0.02\% 
(Logrank test), 0.17\% (Peto \& Peto) and 0.10\% (Peto \& Prentice).  

Figure \ref{rHBLR.fig} shows the
8.4 GHz A--configuration to {\it IRAS} flux ratio distributions of 
sources with known hidden broad--line regions and other type 1 and type 2 
Seyferts in the sample.
Sources with known hidden broad--line regions are significantly more
`radio--loud' than both type 1 Seyferts and type 2 Seyferts not known to 
show hidden broad--line regions. 
The probability of the null hypothesis, that 
the 8.4 GHz A--configuration to {\it IRAS} flux ratio distributions of 
type 1 sources and sources with known 
hidden broad--line regions 
are drawn at random from the same parent population, is: 
3.20\% (Gehan's, perm.), 1.83\% (Gehan's, hyper.), 0.61\% 
(Logrank test), 3.20\% (Peto \& Peto) and 3.49\% (Peto \& Prentice).  
The probability of the null hypothesis, that 
the 8.4 GHz A--configuration to {\it IRAS} flux ratio distributions of 
type 2 sources not known to show hidden broad--line regions 
and type 2 sources with known hidden broad--line regions 
are drawn at random from the same parent population, is: 
1.22\% (Gehan's, perm.), 0.48\% (Gehan's, hyper.), 0.15\% 
(Logrank test), 1.24\% (Peto \& Peto) and 1.28 \% (Peto \& Prentice).  
These results are due to sources with known hidden broad--line regions having
brighter radio fluxes, rather than fainter {\it IRAS} fluxes.
The probability of the null hypothesis, that the 8.4 GHz A--configuration 
flux distributions of known hidden--broad--line sources and 
other type 1 or type 2 sources are drawn at random from the same 
parent population, is:
0.84\% (Gehan's, perm.), 0.18\% (Gehan's, hyper.), 0.12\% 
(Logrank test), 0.83\% (Peto \& Peto) and 0.65\% (Peto \& Prentice).  

Hidden broad--line regions have been
identified in sources throughout the redshift range of the sample.
The probability that the redshift distributions of 
sources with known hidden broad--line regions and all
other Seyferts are drawn at random from the same parent population is:
27.37\% (Gehan's, perm.), 30.00\% (Gehan's, hyper.), 36.75\% 
(Logrank test) and 36.75\% (Peto \& Peto).
There is no evidence that the radio structures in sources with known hidden
broad--line regions differ in size from other Seyferts. 
The probability that the radio sizes of sources with known hidden
broad lines and all other Seyferts are drawn at random from the same
parent population is: 27.93\% (Gehan's, perm.), 25.82\% (Gehan's,
hyper.), 40.18\% (Logrank test) and 33.51\% (Peto \& Peto) 32.38\%
(Peto \& Prentice).

\subsection{Implications: hidden broad--lines are associated with
powerful radio emission.}

Our results provide confirmation that type 2 Seyferts 
with known hidden broad lines have higher radio luminosities
than other type 2 Seyferts, as claimed by \scite{Moran92}. 
\scite{Kay98} noted that Moran et al. did not compare carefully
matched populations of type 1 and type 2 Seyferts, but our study 
suggests that this limitation does not invalidate the earlier result;
we find the same trend using radio luminosity functions
drawn from a single homogeneously--selected sample.
The difference between the radio powers of sources with known
hidden broad--line regions and other Seyferts in the sample
suggests that the observation of hidden broad lines in  
type 2 Seyferts is not simply because they have low--inclination
tori \cite{Heisler97}; torus inclination should not affect
radio power, especially for Seyfert 2s.
In Section \ref{IR.sec} we have suggested that {\it IRAS} flux ratios 
are better indicators of the relative strength of the nuclear 
and disc emission than of the torus inclination, and this interpretation
is supported by the anti--correlation between 
25 $\mu$m to 60 $\mu$m {\it IRAS} flux ratios and morphological type
found for extended 12 $\mu$m Seyferts \cite{Hunt99I} which is hard to
explain via torus inclination.

One explanation of our result 
is that observational limitations prevent the observation of
hidden broad lines in sources with faint radio sources i.e. 
if broad--line luminosities are correlated with 8.4 GHz 
A--configuration luminosities the unidentified population 
of hidden--broad--line sources would have weak radio sources 
\cite{Kay98}.
According to this explanation, hidden broad lines
are only detectable where there is a high contrast between the 
AGN light and the galaxy
light; the high fraction of hidden--broad--line sources among the most
`radio--loud' sources implies that, with sensitive enough observations, 
hidden broad lines should be detectable in around 80\% of all 
type 2 Seyferts.
If {\it IRAS} fluxes contain a mixture of `hot' AGN emission 
and `cooler' starburst/extra--nuclear emission (\S \ref{IR_disc.sec}), 
Seyferts with powerful nuclei are expected to have the hottest  
{\it IRAS} colours, and this may explain 
why Seyferts with hidden broad lines have lower 60 $\mu$m to 25 $\mu$m
{\it IRAS} flux ratios than other type 2 Seyferts \cite{Heisler97}.

An alternative explanation of our result is that hidden--broad--line
sources are intrinsically different from other Seyferts.
The proposition that sources which show hidden broad lines have the
brightest nuclei does not readily explain why they have 
smaller internal extinctions than other type 2 Seyferts \cite{Heisler97}.
In addition, the fact that hidden--broad--line sources have been identified 
throughout the redshift range of the sample suggests that their nuclei 
are not simply the brightest (most nearby) nuclei, but 
the most luminous.
In fact, the identification of hidden broad lines does not always 
require high contrast between the AGN light and the galaxy light.
For example, the optical spectrum of NGC 788 contains a significant 
fraction of starlight \cite{Kay98}; note however, that NGC 788 
has a weak nuclear radio source.

\scite{Moran92} used the excessive radio luminosities of Seyferts with 
hidden broad lines to suggest that there may be two types of 
Seyfert 2 galaxies; hidden type 1 objects and
`true' Seyfert 2 galaxies which lack broad emission lines.
However, this explanation does not explain the fact that
the population of sources with hidden broad lines is more
powerful than the population of type 1 Seyferts.
If the hidden--broad--line galaxies in the sample are well identified, 
they must be different from both `true' type 2
Seyferts and  type 1 Seyferts.

There is no evidence that the difference between 
hidden--broad--line sources and other Seyfert galaxies is related to
radio outflows larger than tens of parsecs because the radio sizes 
and radio morphologies of the two populations are matched.
However, closer to the active nucleus strong radio emission may
indicate regions where electron scatters are abundant, or shock fronts 
which provide `mirrors' for scattering nuclear photons.

\section{SELECTION EFFECTS}
\label{bias.sec}

The selection criteria of the sample studied can be 
the most important sources of systematic error in a statistical study
of Seyfert galaxies. 
The lack of large samples in which
type 1 and type 2 sources are represented equally has been 
an important governing factor for understanding the
generic properties of Seyfert galaxies and has apparently lead to 
misleading comparisons of their radio properties. 
Since it has been selected from the {\it IRAS} catalogue, 
the extended 12 $\mu$m sample is likely to contain Seyferts with 
higher than average star--formation rates, however, it
is one of the largest well--defined samples of Seyfert galaxies
available, and there is no evidence that either the type 1 and 
type 2 sources it contains are not of comparable bolometric luminosity.
Here we attempt to identify its limitations and quantify any 
selection effects which could undermine our results by examining the
selection criteria used.

\subsection{The 12 $\mu$m flux criterion}
\label{12umFlux.sec}

The 12 $\mu$m AGN sample was selected 
according to the premise that an approximately constant fraction of
the bolometric flux of quasars and both types of Seyferts is emitted 
in the {\it IRAS} 12 $\mu$m band \cite{SM89}.
Contrary to this premise, 
models of optically--thick, geometrically--thin dusty tori predict 
that face--on tori (Seyfert 1s according to the unification model) may
emit up to an order of magnitude more of their bolometric
luminosity at mid--infrared wavelengths than edge--on tori
(\pcite{Pier92}; \pcite{Granato94}; \pcite{Efstathiou95}).
Empirical evidence in favour of anisotropic mid--infrared emission
from Seyferts is given by \scite{Heckman95} who 
compared small--aperture 10.6 $\mu$m fluxes with [O {\small III}] fluxes and
low--resolution 1.4 GHz fluxes for four samples of Seyferts 
and found evidence that the mid--infrared luminosities of type 1
Seyfert  galaxies are 2 to 4 times higher than in type 2 Seyfert galaxies. 
According to this result, the extended 12 $\mu$m Seyfert sample is
expected to be biased against low--luminosity type 2 Seyferts
because of the {\it IRAS} 12 $\mu$m flux limit.
In order to test this prediction we have compared the 
redshift distributions and 12 $\mu$m flux distributions 
of type 1 and type 2 Seyferts in the sample.

\begin{figure} 
\centerline{       
\includegraphics[angle=0,width=8.5cm]{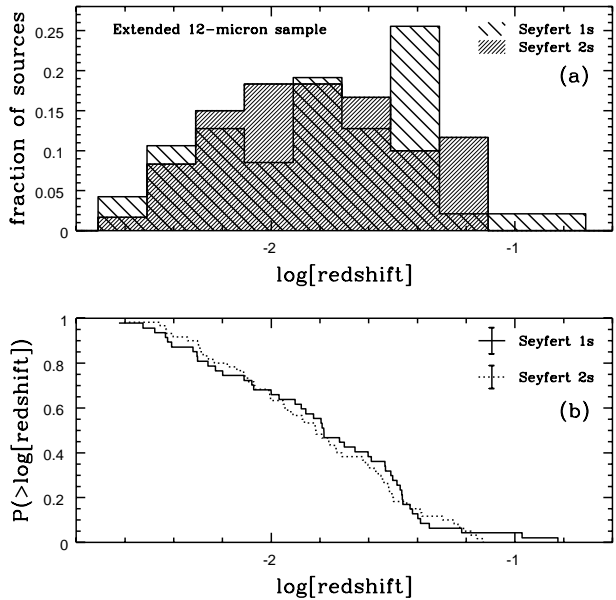}}
\caption{(a) Histograms showing the fractional redshift 
distributions of the 47 type 1 Seyferts and the 60 type 2 Seyferts 
in the extended 12 $\mu$m Seyfert sample. 
(b) The cumulative redshift distributions. 
The fact that the two redshift distributions are likely to be drawn 
from the same parent population provides evidence that the sample has 
been selected in a way which will not bias the comparison of type 1
and type 2 sources.  
}
\label{12u_z.fig}
\end{figure}

\begin{figure} 
\centerline{       
\includegraphics[angle=0,width=8.5cm]{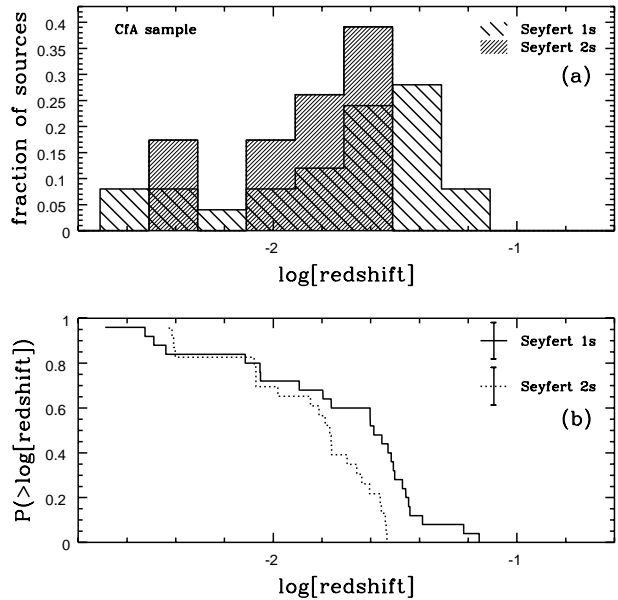}}
\caption{The fractional redshift distributions of the 26 type 1 
Seyferts and the 22 type 2 Seyferts in the CfA sample 
indicate a possible bias against the
selection of distant Seyfert 2s.}
\label{CfA_z.fig}
\end{figure}

The redshift distributions of different classes of object in a
flux--limited sample provide indicators of whether they are equally
luminous at the wavelength of  
selection; since the observed flux of an object of given luminosity 
falls proportionally with the square of its distance,
differences between the luminosities of two classes of source at the
wavelength of selection can easily produce mismatched redshift
distributions.
Therefore, if type 2 Seyferts are truly less luminous than type 1
Seyferts in the {\it IRAS} 12 $\mu$m band, we expect them to be drawn from a
smaller volume of space than type 1 Seyferts in the extended 12 $\mu$m sample.

Figure \ref{12u_z.fig} shows the redshift distributions of the two
Seyfert types in the extended 12 $\mu$m sample.
The two types are equally distributed according to redshift. 
The probability of the null hypothesis, that the redshift distributions
of the two Seyfert types are drawn at random from the same parent population, 
is: 94.48\% (KS test), 83.84\% (Gehan's perm.), 83.86\% (Gehan's 
hyper.), 68.70\% (Logrank test) and 68.70\% (Peto \& Peto). 
The mean redshift of Seyfert 1s is 0.024 $\pm$ 0.004       
and that of Seyfert 2s is 0.022 $\pm$ 0.002: the larger difference 
between the mean redshifts of type 1 and type 2 objects in the
extended 12 $\mu$m AGN sample noted by \scite{RMS93} and \scite{Hunt99I} 
is due to the inclusion of the 5 radio--loud  sources we have
excluded in Paper I \cite{Thean99}. 
The redshift distributions show no evidence of 
the identification bias against faint Seyfert 2s discussed
in Section \ref{ID.sec};
note that a bias against faint type 2 Seyferts 
 acts in the opposite sense to that required to
reconcile the observed redshift and 12 $\mu$m 
flux density distributions with the results of \scite{Heckman95}. 

For comparison with the extended 12 $\mu$m Seyfert sample, Figure 
\ref{CfA_z.fig} shows the redshift distributions of the CfA Seyfert sample 
\cite{Huchra92}. 
The CfA Seyfert sample has the advantage of being selected 
spectroscopically without the use of (biased) catalogues of AGN, however, the 
redshift distributions the two Seyfert types in this sample appear to differ. 
The probability of the null hypothesis, that the two Seyfert types 
are drawn at random from the same parent population, is: 
5.9\% (KS test) 8.4\% (Gehan's perm.), 7.9\% (Gehan's hyper.), 0.4\% (Logrank
test)and 0.4\% (Peto \& Peto). 
These test results are inconclusive, but the sense of the difference 
in the two populations and the results of the Logrank and Peto \& Peto
tests are suggestive of a bias against distant type 2 Seyferts. 
This type of bias could be explained by the criteria used to define the CfA
sample; at the wavelength of selection Seyfert type 1 nuclei are more
luminous than type 2 nuclei and can account for a significant fraction of
the light of the host galaxy, in addition, broad emission line profiles
make distant type 1 sources easier to identify \cite{Kukula95}. 

\begin{figure} 
\centerline{      
\includegraphics[angle=0,width=8.5cm]{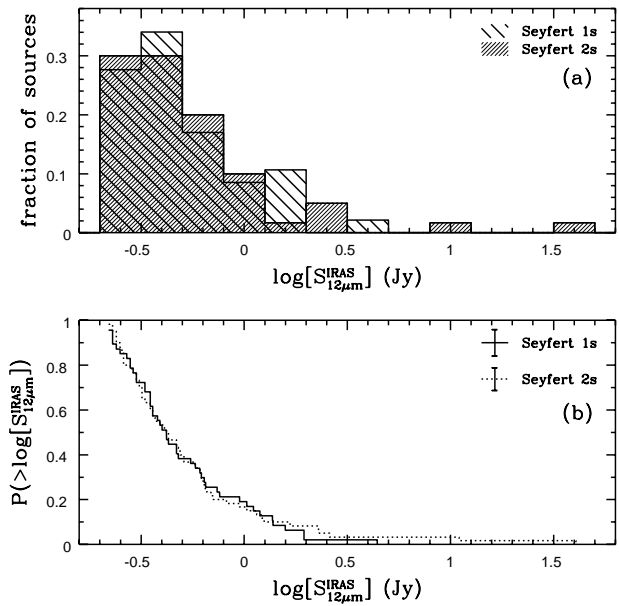}}
\caption{(a) Histograms showing the fractional {\it IRAS} 
12 $\mu$m flux density distributions of the  47 type 1 Seyferts and the 60 
type 2 Seyferts in the extended
12 $\mu$m sample. (b) The cumulative 12 $\mu$m flux density distributions. 
The flux density distributions are well matched and together with the redshift 
distributions, this implies that the two Seyfert types are equally
luminous at the wavelength of selection. 
} 
\label{S12.fig}
\end{figure}

From their matched redshift distributions, we have inferred that
type 1 and type 2 Seyferts in the extended 12 $\mu$m sample are
equally luminous in the {\it IRAS} 12 $\mu$m band (8.5 $\mu$m to 15 $\mu$m). 
This is supported by the fact the {\it IRAS} 12 $\mu$m flux density 
distributions of type 1 and type 2 Seyferts in the sample are also
well matched (see Fig. \ref{S12.fig}).  
The probability of the null hypothesis, that the {\it IRAS} 12 $\mu$m
fluxes of type 1 and type 2 Seyferts are drawn at random from the same
parent population, is:
98.31\% (KS test), 100.00\% (Gehan's perm.),  100.00\% (Gehan's
hyper.), 76.12\% (Logrank) and 76.12\% (Peto \& Peto).

In summary, our results suggest that the {\it IRAS} 12 $\mu$m 
emission from type 1 and type 2 Seyferts is of roughly equal strength.
They are in good agreement with those of
\scite{Fadda98} who argued against early models of 
compact, very optically--thick tori.
A possible explanation for the low  
small--aperture 10.6 $\mu$m to 1.4 GHz single--dish flux ratios 
of type 2 Seyferts in the extended 12 $\mu$m sample
\cite{Heckman95} is that type 2 Seyferts have brighter non--nuclear
radio fluxes (\S \ref{star2.sec});
note that the small--aperture 10.6 $\mu$m 
to [O{\small III}] flux ratios of type 1 and type 2 Seyferts from 
the extended 12 $\mu$m sample are indistinguishable.

\subsection{Infrared colour criteria}
\label{IrCol.sec}

In Section \ref{S1.4-IR12.sec} we showed that the {\it IRAS} 60 $\mu$m flux 
distributions of the type 1 and type 2 Seyferts in the extended 12
$\mu$m sample are different, and this means that the colour 
criteria used to define the sample could be a source of bias.
The mid-- to far--infrared colour distributions are not 
significantly truncated by the explicit infrared 
colour criterion used to define the sample (S$\rm _{12\mu m}$/S$\rm _{60\mu
m}\leq$ 2 or S$\rm _{12\mu m}$/S$\rm _{100\mu m}\leq$ 1) and these are
 unlikely to cause a significant bias.
However, a further selection criterion, which was used in 
order to ensure real {\it IRAS} detections, requires all sources to have   
moderate or high flux quality flags on 60 $\mu$m or 100 $\mu$m fluxes 
as well as on 12 $\mu$m fluxes. 
This acts as a `hidden colour criterion' and tends to bias 
the sample against faint sources with high mid-- to far--infrared flux
ratios i.e. the {\it IRAS} sensitivity at both 60 $\mu$m and 100
$\mu$m is lower than that at 12 $\mu$m so, in order to be selected,
sources near the 12 $\mu$m flux  limit must have lower mid-- to
far--infrared flux ratios than required by the explicit infrared
colour criterion.  
Taking 3--$\sigma$ detection thresholds\footnote{Average noise values
were taken from the IRAS website 
(www.ipac.caltech.edu/ipac/iras/iras\_mission.html).} at 
12 $\mu$m, 60 $\mu$m and 100 $\mu$m  of 0.21 Jy, 0.26 Jy and 0.90 Jy 
respectively,
the `hidden colour criterion' requires that sources at the flux limit
must satisfy the criterion S$\rm _{12\mu m}$/S$\rm _{60\mu m} \leq$ 0.85 or
S$\rm _{12\mu m}$/S$\rm _{100\mu m} \leq$ 0.24.  
Since type 1 Seyferts are more likely to have high mid-- to
far--infrared flux ratios than type 2 Seyferts,
the hidden colour criterion can 
result in a bias against weak type 1 Seyferts.
For sources with {\it IRAS} 12 $\mu$m flux densities brighter than 0.52 Jy
($\sim$40\% of the sample) the {\it IRAS} 60 $\mu$m sensitivity is
enough to detect any sources which would satisfy the S$\rm _{12\mu
m}$/S$\rm _{60\mu m}$ $\leq$ 2 condition (no sources in the AGN sample fail
the S$\rm _{12\mu m}$/S$\rm _{60\mu m}$ $\leq$ 2 condition and are
selected on the basis of their 100 $\mu$m flux density). 
Of these bright sources, only two do not satisfy the `hidden colour
criterion', and both are type 1 Seyferts (F05563-3820 and MK704). 
If a similar fraction of faint sources are missing from 
the full sample, around 3 faint Seyferts (most likely type 1s) which
have the 12 $\mu$m flux densities above the flux limit will have been omitted
because they have high mid-- to far--infrared flux ratios.

\subsection{The AGN identification procedure}
\label{ID.sec}

\begin{figure} 
\centerline{      
\includegraphics[angle=0,width=8.5cm]{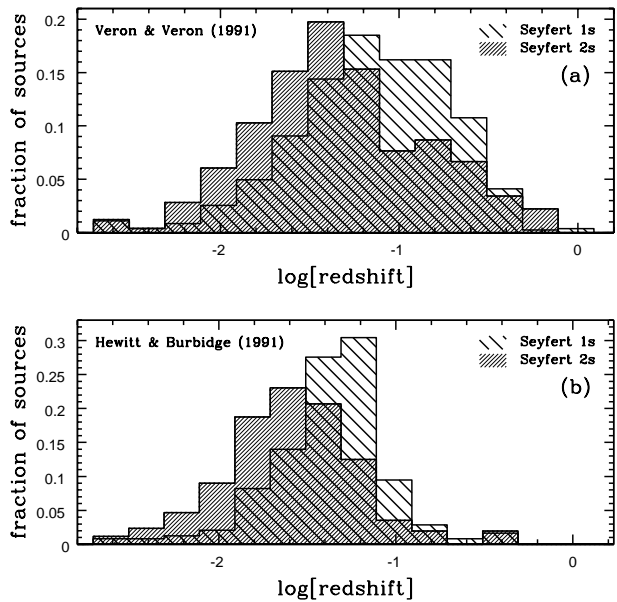}}
\caption{
Histograms showing the fractional redshift distributions of
the two Seyfert types in (a) Veron \& Veron (1991) and (b) Hewitt \&
Burbidge (1991), two of the catalogues used to identify AGN in the
extended 12 $\mu$m sample.  
The relative fraction of Seyfert 1s in these catalogues increases
steadily with redshift indicating a bias against distant Seyfert type
2 sources.  
There are a total of 827 Seyfert 1s and 496 Seyfert 2s
listed in Veron \& Veron (1991) and 243 Seyfert 1s and 256 Seyfert 2s
listed in Hewitt \& Burbidge (1991).
} 
\label{cats.fig}
\end{figure}

One limitation of the extended 12 $\mu$m AGN sample is that 
AGN were identified by cross--referencing with the literature
without a complete and homogeneous spectroscopic identification
procedure. 
Although redshifts are available for 98\% of the extended
12 $\mu$m galaxy sample, the quality of the spectroscopy
used to provide them is insufficient to identify Seyferts with weak
emission lines.  
For this reason, Seyferts in the extended 12 $\mu$m sample can be
difficult to identify.
A number of sources were not recognised as AGN at the time 
\scite{RMS93} defined the AGN sample. 
By cross--referencing the extended 12 $\mu$m sample with the NED, 
\scite{Hunt99I} found 29 galaxies which were classified as 
`normal galaxies' or `high--far--infrared' galaxies by Rush et al.,
but which are classified as Seyferts in the NED.
We have identified 38 such
sources by the same procedure (see Appendix \ref{newSy.sec});
note that Hunt et al. also found that 22 of Seyferts identified 
by Rush et al. are not classified as Seyferts in the NED.

From V/V$\rm _{max}$ completeness tests, Rush et al. initially 
suspected that some weak type 1 Seyferts remained to be identified in the 
sample, however,  nearly all of the newly--recognised Seyferts are
type 2 sources.
This result is in the sense predicted by \scite{Barcons95}
who estimated that up to 20\% of the weak Seyfert 1s and
50\% of the  weak Seyfert 2s were missing from the sample.
The most likely reasons that the majority of the newly--recognised
Seyferts are type 2s is that AGN search techniques can 
favour the discovery of Seyfert 1s and Seyfert 2s with faint emission
lines are hard to identify with low quality optical spectra.
Figure \ref{cats.fig} shows 
the redshift distributions of type 1 and
type 2 Seyferts from two of the AGN catalogues used to identify AGN in the 
extended 12 $\mu$m sample. 
A bias against distant (faint) type 2 Seyferts  
is evident in both catalogues.
The ratio of the number of Seyfert 1s to the number 
of Seyfert 2s contained in these catalogues increases steadily 
with redshift and there is an extremely low probability that the 
redshift distributions of type 1
and type 2 Seyferts are drawn at random from the same parent population: 
2$\times$10$^{-14}$\% (KS) for \scite{Veron-Cetty91} and 
8$\times$10$^{-10}$\% (KS) for \scite{Hewitt91}. 

The radio properties of the newly--identified type 2 Seyferts in the
sample may affect several of our results, particularly if
they are low--luminosity sources.
For example, correlations between radio power and
forbidden--line luminosities (\pcite{deBruyn+W78};
\pcite{Whittle92III}) 
may mean that we have over--estimated the
strengths of type 2 Seyferts. 
The complete identification of Seyferts from the extended
12 $\mu$m galaxy sample and follow--up radio observations 
would be useful for improving the accuracy of our analysis.

\section{SUMMARY}
\label{summary.sec}

The main results of our analysis may be summarised as follows:

\begin{itemize}

\item
Eighty--five of the 98 Seyferts from the extended 12 $\mu$m AGN sample 
observed were detected by the VLA in A--configuration at 8.4 GHz. 
They showed a range of radio morphologies; roughly 50\% were
unresolved, 20\% were slightly--resolved, 10\% had two components,
10\% had two or more linearly--aligned components and 10\% had diffuse
or ambiguous morphologies.

\item
There is no significant difference between the luminosities of compact 
radio components in type 1 and type 2 Seyferts.
This result is consistent with the Seyfert unification model and
suggests that the central engines of each type of Seyfert
are similar. 

\item
There is no significant difference between the sizes or morphologies 
of compact 
radio components in type 1 and type 2 Seyferts in the sample.
This result is consistent with the Seyfert unification model and
suggests that each type of Seyfert has a similar capacity to produce 
extended radio outflows (our comparison is probably 
insensitive to the effects of orientation). 

\item
There is a significant correlation between radio power and radio 
size for type 2 Seyferts which show evidence of collimated nuclear radio 
sources.
The same correlation is not shown by type 1 Seyferts.

\item
The {\it IRAS} emission from infrared--bright 
Seyfert galaxies is probably dominated by kiloparsec--scale
emission regions, possibly star--forming regions.

\item 
There is evidence of a correlation between host galaxy morphology and
nuclear radio power, with early--type galaxies showing more luminous
radio cores.

\item 
Collimated nuclear radio sources appear to be randomly oriented with
respect to the galactic plane.

\item 
The nuclear radio sources of Seyferts with known 
V--shaped extended emission--line
regions, indicative of 'ionisation cones', 
are larger than other sources in the sample.

\item 
The nuclear radio sources of Seyferts with known hidden broad--line
regions are more powerful than other types of Seyfert.

\item
The extended 12 $\mu$m sample appears to contain matched 
populations of type 1 and type 2 Seyfert galaxies.

\end{itemize}

\section{ACKNOWLEDGMENTS}

AHCT would like to acknowledge the receipt of a studentship from
the Particle Physics and Astronomy Research Council and a visit funded
by the STScI visitor program. Part of this research was 
supported by the European Commission, TMR Programme, Research Network Contract
ERBFMRXCT96-0034 ``CERES''. 
We have made use of NASA's Astrophysics Data
System Abstract Service, the NASA/IPAC Extragalactic database (NED),
which is operated by the Jet Propulsion Laboratory, and 
{\small ASURV} Rev 1.1 \cite{Lavalley92}, which implements
the methods presented in \scite{Feigelson+N85} and \scite{Isobe86}. 
We thank the referee, R. Antonucci, for helpful comments.

\appendix

\section{Archive data}
\label{archive.sec}

\subsection{Host galaxy morphologies}

Morphological classifications were available in the LEDA for 78
galaxies from the extended 12 $\mu$m Seyfert sample and are given in 
Table \ref{morph.tab} which is arranged as follows;
{\it Column 1:} Galaxy name;
{\it Column 2:} Galaxy type, T$_{\rm Sy}$. 
{\it Column 3:} Hubble type, T$\rm _{Hubb}$, as
obtained from the NED;
{\it Column 4:} Morphological type code, T, and its associated
uncertainty, $\Delta$T, as obtained from the LEDA. 
Galaxy morphological types in the LEDA are codified using the 
continuous parameter T following the methods outlined in
The Third Reference Catalogue of Bright Galaxies (RC3) 
\cite{deVaucouleurs91RC3},  
low values of T correspond to early--type galaxies 
(Fig. \ref{t-P.fig} shows the ranges in T which correspond to
the various Hubble types). 
{\it Column 5:} Position angle of the galaxy major axis, PA, in degrees; 
{\it Column 6:} Inclination of the galaxy, i, in degrees. 

\begin{table}
\tiny
\begin{center}
\begin{tabular}{|l|c|l|r|r|r|} \hline 
{\bf Galaxy} & {\bf T$_{\rm Sy}$} & {\bf T$_{\rm Hubb}$} & {\bf T$\pm\Delta$T}~~ & {\bf PA} & {\bf i~} \\
 & &  & & {\bf ($^{\circ}$)} & {\bf ($^{\circ}$)} \\
\hline
        ~NGC262=Mrk348   & 2 &     S0a &  $-$0.472$\pm$0.935 &   -- &   14  \\
        ~NGC424=TOL0109 & 2 &  S0a--R &  0.077$\pm$0.561 &  60 &   63  \\
        ~NGC526A        & 1 &      S0 & $-$1.906$\pm$0.967 &   -- &   58  \\
        ~NGC513         & 2 &      Sc &  6.200$\pm$0.800 &  75 &   61  \\
        ~Mrk1034       & 1 &     Sab &  2.320$\pm$2.557 & 117 &    0  \\
        ~MCG--3--7--11     & 1 &      Sc &  5.015$\pm$0.469 &   2 &   42  \\
        ~NGC931=Mrk1040  & 1 &     Sbc &  3.651$\pm$0.608 &   -- &   75  \\
        ~NGC1056=Mrk1183 & 2 &      Sa &  1.000$\pm$0.384 & 160 &   56  \\
        ~NGC1068        & 2 &      Sb &  2.985$\pm$0.300 &  70 &   31  \\
        ~NGC1125        & 2 &    S0a &  0.024$\pm$0.506 &   -- &   60  \\
        ~NGC1143/4      & 2 &       E & $-$4.195$\pm$2.553 & 130 &   48  \\
        ~MCG--2--8--39     & 2 &     SBa &  1.000$\pm$0.500 &  10 &   50  \\
        ~NGC1194        & 2 &    S0a &  $-$0.730$\pm$0.556 & 140 &   51  \\
        ~NGC1241        & 2 &     SBb &  3.048$\pm$0.389 & 145 &   49  \\
        ~NGC1320=Mrk607  & 2 &      Sa &  1.053$\pm$0.641 &  45 &   70  \\
        ~NGC1365        & 1 &     SBb &  3.165$\pm$0.615 &  32 &   55  \\
$^{\ast}$NGC1566        & 1 &    SBbc &  4.008$\pm$0.142 &  60 &   42  \\
        ~Mrk618        & 1 &     SBb &  2.989$\pm$0.494 &  85 &   41  \\
        ~NGC1667        & 2 &     SBc &  4.996$\pm$0.442 &  20 &   39  \\
$^{\ast}$E33--G2       & 2 &      S0 & $-$2.200$\pm$1.800 &  10 &   20  \\
        ~MCG--5--13--17    & 1 &  S0a--M &  0.086$\pm$0.597 & 160 &   53  \\
        ~E253--G3      & 2 &      S? &  1.600$\pm$4.800 & 108 &   44  \\
        ~Mrk6          & 2 &    S0a &  $-$0.693$\pm$0.712 & 130 &   51  \\
        ~Mrk9          & 1 &      S0 & $-$2.000$\pm$0.500 &   -- &   35  \\
        ~Mrk79         & 1 &     SBb &  3.034$\pm$0.386 &   -- &    4  \\
        ~NGC2639        & 1 &      Sa &  0.997$\pm$0.412 & 140 &   40  \\
        ~NGC2992        & 2 &    Sa--M &  0.930$\pm$0.481 &  15 &   76  \\
        ~NGC3079        & 2 &     SBc &  6.447$\pm$1.055 & 165 &   81  \\
        ~NGC3227        & 1 &   SBa--M &  1.472$\pm$0.842 & 155 &   46  \\
        ~NGC3511        & 1 &   SBc--M &  5.065$\pm$0.393 &  76 &   70  \\
        ~NGC3516        & 1 &    S0--R & $-$1.977$\pm$0.396 &   -- &   34  \\
        ~M+0--29--23    & 2 &      Sb &  2.989$\pm$0.494 &   -- &   32  \\
        ~NGC3660        & 2 &  SBbc--R &  3.852$\pm$0.487 & 115 &   36  \\
        ~NGC3982        & 2 &     SBb &  3.118$\pm$0.525 &   -- &   29  \\
        ~NGC4051        & 1 &    SBbc &  4.002$\pm$0.142 & 135 &   34  \\
        ~UGC7064        & 1 &     SBb &  3.226$\pm$0.682 &   -- &    0  \\
        ~NGC4151        & 1 &  SBab--R &  2.025$\pm$0.397 &  50 &   38  \\
        ~Mrk766        & 1 &     SBa &  0.958$\pm$0.670 &   -- &   21  \\
        ~NGC4388        & 2 &      Sb &  2.773$\pm$0.519 &  92 &   74  \\
        ~NGC4501        & 2 &      Sb &  3.440$\pm$0.630 & 140 &   57  \\
        ~NGC4579        & 1 &     SBb &  2.807$\pm$0.421 &  95 &   37  \\
        ~NGC4593        & 1 &   SBb--M &  3.026$\pm$0.382 &  55 &   45  \\
        ~NGC4594        & 1 &      Sa &  1.101$\pm$0.330 &  89 &   59  \\
        ~NGC4602        & 1 &    SBbc &  4.020$\pm$0.236 &  81 &   70  \\
        ~Mrk231=UGC8058  & 1 &      Sc &  4.988$\pm$0.462 &  10 &   42  \\
        ~NGC4922A/B     & 2 &     E--M & $-$4.370$\pm$1.482 &   -- &   14  \\
        ~NGC4941        & 2 &    SBab &  2.365$\pm$0.731 &  15 &   51  \\
        ~NGC4968        & 2 &      S0 & $-$2.010$\pm$0.399 &  56 &   61  \\
        ~NGC5005        & 2 &    SBbc &  3.965$\pm$0.194 &  65 &   60  \\
        ~NGC5033        & 1 &      Sc &  5.120$\pm$0.658 & 170 &   67  \\
        ~MCG--3--34--63    & 2 &      S? &  2.700$\pm$5.200 &   -- &   72  \\
        ~NGC5135        & 2 &    SBab &  2.316$\pm$0.597 &   -- &   21  \\
        ~NGC5194=M51    & 2 &   Sbc--M &  4.041$\pm$0.290 & 163 &   46  \\
        ~MCG--6--30--15    & 1 &    E--S0 & $-$0.767 $\pm$3.242 & 116 &   54  \\
        ~NGC5256=Mrk266  & 2 &     Sab &  1.700$\pm$2.500 &   -- &   32  \\
        ~I4329A       & 1 &    S0a & $-$1.051$\pm$0.480 &  45 &   72  \\
        ~NGC5347        & 2 &  SBab--R &  2.029$\pm$0.336 & 130 &   38  \\
        ~Mrk463        & 2 &      S? &  2.700$\pm$6.800 &   -- &   60  \\
        ~NGC5506        & 2 &      Sa &  1.326$\pm$1.266 &  91 &   74  \\
        ~NGC5548        & 1 &  S0a--R &  0.428$\pm$0.675 & 110 &   34  \\
        ~NGC5929        & 2 &    Sa--M &  1.362$\pm$1.808 &   -- &   16  \\
        ~NGC5953        & 2 &  S0a--M &  0.236$\pm$1.309 & 169 &   43  \\
        ~MCG--2--40--4     & 2 &      Sa &  1.000$\pm$0.500 & 120 &   38  \\
$^{\ast}$E141--G55     & 1 &      Sc &  5.200$\pm$3.000 &  50 &   48  \\
$^{\ast}$NGC6860        & 1 &      Sb &  2.961$\pm$0.463 &  34 &   56  \\
        ~NGC6890        & 2 &     SBb &  2.811$\pm$0.587 & 152 &   36  \\
$^{\ast}$I5063        & 2 &    S0a & $-$1.019$\pm$0.660 & 116 &   48  \\
        ~UGC11680=Mrk897 & 2 &    Sc--M &  5.856$\pm$0.788 &  70 &   48  \\
        ~NGC7130=I5135  & 2 &      Sa &  1.065$\pm$0.723 &   -- &   22  \\
        ~NGC7172        & 2 &      Sa &  1.050$\pm$0.685 & 100 &   56  \\
        ~NGC7213        & 1 &      Sa &  0.924$\pm$0.624 &   -- &   28  \\
        ~NGC7314        & 1 &    SBbc &  4.017$\pm$0.164 &   3 &   64  \\
        ~NGC7469        & 1 &     SBa &  1.113$\pm$0.321 & 125 &   45  \\
        ~NGC7496        & 2 &     SBb &  3.159$\pm$0.647 &   -- &   18  \\
        ~NGC7582        & 2 &    SBab &  2.020$\pm$0.485 & 157 &   63  \\
        ~NGC7590        & 2 &     Sbc &  4.003$\pm$0.328 &  36 &   67  \\
        ~NGC7603=Mrk530  & 1 &      Sb &  3.018$\pm$0.471 & 165 &   49  \\
        ~NGC7674=Mrk533  & 2 &  SBbc--M &  3.796$\pm$0.561 &   -- &   24  \\
\hline
\end{tabular}
\caption{Morphological information for the host galaxies of the 
extended 12 $\mu$m Seyfert sample.}
\label{morph.tab}
\end{center}
\end{table}

\subsection{NVSS components}

Table \ref{NVSS.tab} gives the VLA D and DnC configuration 
1.4 GHz radio flux densities obtained from the NVSS for 92 of the 113 
radio--quiet sources from the extended 12 $\mu$m sample.
NVSS counterparts within 10 arcsec of the A--configuration
positions are found for 85 of the 87 sources at declinations north of
$-$40 degrees for which A--configuration positions are available 
(there is less than a 0.2\% probability of finding a NVSS source within
10 arcsec of an arbitrary position in the sky). 
Seven out of ten of the sources north of declination $-$40 degrees and
without A--configuration positions have NVSS counterparts within 20
arcsec of the IRAS position (there is a 0.5\% probability of
finding a NVSS source within 20 arcsec of an arbitrary position in the
sky). 
Sixteen sources are south of $-$40 degrees declination in a region of
sky not covered by the NVSS.
Table \ref{NVSS.tab} is arranged as follows;
{\it Column 1:} Galaxy name;
{\it Column 2:} Galaxy type, starbursts (s) and LINERs (L)
have been excluded from our analysis;
{\it Column 3 and 4:} Right ascension and declination of the NVSS
components determined from Gaussian fits;
{\it Column 5:} The separation of the NVSS component
position and the VLA 8.4 GHz A--configuration or {\it IRAS} position,
$\rm \theta_{sep}$, in arcseconds.
The rms uncertainties in right ascension and declination for the NVSS are
0.3 arcsec for strong point sources (S $>$ 30 mJy/beam) and 5 arcsec for
the faintest sources detectable (S = 2.5 mJy/beam). 
Sources whose NVSS counterparts were found by searching on {\it IRAS} positions rather
than 8.4 GHz VLA A--configuration positions are denoted by
an asterisk symbol ($\ast$);
{\it Column 6:} Peak NVSS flux density determined from a Gaussian fit 
(mJy/beam).
NVSS flux density upper--limits of 5--$\sigma$ have been assumed for 5 sources
for which no NVSS counterpart was found within the search radius
(1--$\sigma$ = 0.45 mJy/beam);
{\it Column 7:} Integrated NVSS flux density (mJy/beam). 

\begin{table*}
\tiny
\begin{center}
\begin{tabular}{|l|c|r|r|r|r|r|} \hline 
{\bf Galaxy} & {\bf Type} & {\bf RA(J2000)} & {\bf Dec(J2000)} & {\bf
$\rm \theta_{sep}$} & {\bf S$\rm ^{NVSS}_{peak}$} & {\bf S$\rm ^{NVSS}_{integ.}$} \\
 &  & {\bf (h m s)~~~} & {\bf ($^{\circ}$ $'$ $''$)~~} & {\bf
($''$)} & {\bf (mJy/B)} & {\bf (mJy/B)} \\
\hline
Mrk 335        & 1 & 00 06 19.45 & +20 12  10.3 &          1.4 &    7.6 &   7.6 \\
NGC 34=Mrk 938    & s & 00 11 06.54 & $-$12 06  27.4 &          0.4 &   62.4 &  67.8 \\
NGC 262=Mrk 348   & 2 & 00 48 47.15 & +31 57  25.3 &          0.3 &  292.7 & 292.7 \\
IZW1         & 1 & 00 53 34.77 & +12 41  33.8 &          3.3 &    8.8 &   8.8 \\
E541-IG12    & 2 & 01 02 17.91 & $-$19 40  06.5 &          7.7 &    4.8 &   4.8 \\
NGC 424=TOL0109 & 2 & 01 11 27.71 & $-$38 04  59.6 &          1.3 &   23.3 &  23.3 \\
NGC 526A        & 1 & 01 23 54.57 & $-$35 03  57.5 &          2.9 &   11.5 &  13.9 \\
NGC 513         & 2 & 01 24 26.66 & +33 47  55.8 &          3.0 &   48.4 &  53.8 \\
F01475-0740  & 2 & 01 50 02.66 & $-$07 25  49.2 &          0.8 &  318.8 & 318.8 \\
Mrk 1034       & 1 & 02 23 21.71 & +32 11  47.0 &          3.7 &   42.3 &  51.6 \\
MCG-3-7-11     & 1 & 02 24 40.45 & $-$19 08  31.7 &          1.8 &   28.2 &  31.3 \\
NGC 931=Mrk 1040  & 1 & 02 25 16.46 & +31 05  19.9 & $^{\ast}$7.1 &    9.3 &  15.5 \\
NGC 1056=Mrk 1183 & 2 & 02 42 48.28 & +28 34  28.1 &          3.6 &   27.8 &  38.6 \\
NGC 1068        & 2 & 02 42 40.72 & $-$00 00  47.7 &          4.2 & 4468.5 & 4857.1 \\
NGC 1097        & L & 02 46 18.93 & $-$30 16  29.3 &          0.9 &  200.6 & 250.7 \\
NGC 1125        & 2 & 02 51 40.44 & $-$16 39  02.1 &          0.2 &   54.8 &  58.3 \\
NGC 1143/4      & 2 & 02 55 12.16 & $-$00 10  58.7 &          2.3 &  140.1 & 155.7 \\
MCG-2-8-39     & 2 & 02 58 06.76 & $-$11 36  50.2 & $^{\ast}$7.5 &    9.0 &   9.0 \\
NGC 1194        & 2 & 03 03 49.67 & $-$01 06  17.3 &          9.1 &    2.9 &   2.9 \\
NGC 1320=Mrk 607  & 2 & 03 24 48.81 & $-$03 02  29.5 &          3.5 &    6.5 &   6.5 \\
NGC 1365        & 1 & 03 33 36.46 & $-$36 08  25.9 &          5.6 &  331.1 & 376.2 \\
NGC 1386        & s & 03 36 46.22 & $-$35 59  57.3 &          0.3 &   35.1 &  37.7 \\
F03362-1642  & 2 & 03 38 33.58 & $-$16 32  15.6 &          2.8 &    9.3 &   9.3 \\
F03450+0055  & 1 & 03 47 40.17 & +01 05  14.6 &          0.7 &   32.5 &  32.5 \\
Mrk 618        & 1 & 04 36 22.34 & $-$10 22  32.7 &          1.4 &   17.3 &  17.3 \\
F04385-0828  & 2 & 04 40 54.33 & $-$08 22  20.7 &          9.5 &   11.7 &  18.9 \\
NGC 1667        & 2 & 04 48 37.06 & $-$06 19  13.0 &          1.9 &   56.9 &  76.9 \\
MCG-5-13-17    & 1 & 05 19 35.72 & $-$32 39  29.5 &          1.5 &   12.7 &  14.9 \\
F05189-2524  & 2 & 05 21 01.38 & $-$25 21  45.0 &          0.4 &   29.1 &  29.1 \\
F05563-3820  & 1 & 05 58 02.20 & $-$38 20  02.4 &          5.9 &   34.9 &  34.9 \\
Mrk 6          & 2 & 06 52 12.47 & +74 25  37.0 &          0.6 &  259.7 & 276.3 \\
Mrk 9          & 1 & 07 36 56.17 & +58 46  16.9 &          7.4 &    4.1 &   4.1 \\
Mrk 79         & 1 & 07 42 32.84 & +49 48  34.1 &          0.8 &   18.3 &  22.3 \\
F07599+6508  & 1 & 08 04 30.75 & +64 59  53.5 &          1.9 &   39.7 &  39.7 \\
NGC 2639        & 1 & 08 43 38.16 & +50 12  20.4 &          0.9 &  109.0 & 116.0 \\
F08572+3915  & 2 & 09 00 25.58 & +39 03  52.8 &          2.7 &    4.7 &   4.7 \\
Mrk 704        & 1 & 09 18 26.05 & +16 18  21.2 &          1.7 &    6.6 &   6.6 \\
UGC 5101        & 1 & 09 35 51.83 & +61 21  12.9 &          2.0 &  145.1 & 170.7 \\
NGC 2992        & 2 & 09 45 42.00 & $-$14 19  34.6 &          0.8 &  226.6 & 226.6 \\
Mrk 1239       & 1 & 09 52 19.11 & $-$01 36  43.5 &          0.2 &   62.8 &  62.8 \\
NGC 3031=M 81    & L & 09 55 33.28 & +69 03  55.0 &          0.6 &   82.4 &  88.6 \\
NGC 3079        & 2 & 10 01 57.82 & +55 40  48.5 &          1.2 &  491.9 & 768.6 \\
NGC 3227        & 1 & 10 23 30.55 & +19 51  54.8 &          0.7 &   86.2 & 100.2 \\
NGC 3511        & 1 & 11 00 57.39 & $-$22 49  03.9 & $^{\ast}$1.5 &   15.4 &  77.0 \\
NGC 3516        & 1 & 11 06 47.71 & +72 34  10.3 &          3.2 &   27.6 &  32.5 \\
M+0-29-23    & 2 & 11 18 38.95 & $-$02 42  34.7 & $^{\ast}$18.3 &   35.3 &  35.3 \\
NGC 3660        & 2 & 11 21 00.18 & $-$08 22  56.1 & $^{\ast}$12.6 &    5.7 &  14.6 \\
NGC 3982        & 2 & 11 56 27.95 & +55 07  30.9 &          1.6 &   37.9 &  57.4 \\
NGC 4051        & 1 & 12 03 09.28 & +44 31  54.0 &          3.7 &   19.6 &  98.0 \\
NGC 4151        & 1 & 12 10 32.52 & +39 24  20.9 &          1.1 &  360.1 & 360.1 \\
Mrk 766        & 1 & 12 18 26.43 & +29 48  47.3 &          1.3 &   37.0 &  39.8 \\
NGC 4388        & 2 & 12 25 46.97 & +12 39  43.7 &          3.8 &   94.5 & 121.2 \\
NGC 4579        & 1 & 12 37 43.74 & +11 49  08.0 &          4.2 &   58.0 &  98.3 \\
NGC 4593        & 1 & 12 39 39.31 & $-$05 20  35.9 &          3.7 &    4.8 &   4.8 \\
NGC 4594        & 1 & 12 39 59.40 & $-$11 37  23.5 &          0.6 &   85.0 &  94.4 \\
NGC 4602        & 1 & 12 38 02.13 & $-$04 51  25.3 & $^{\ast}$18.0 &   16.3 &  40.8 \\
TOL1238-364  & 2 & 12 40 52.85 & $-$36 45  20.1 &          1.0 &   73.6 &  88.7 \\
MCG-2-33-34    & 1 & 12 52 12.65 & $-$13 24  52.8 &          2.6 &   14.3 &  14.3 \\
Mrk 231=UGC 8058  & 1 & 12 56 14.13 & +56 52  23.8 &          1.7 &  271.9 & 309.0 \\
NGC 4922A/B     & 2 & 13 01 25.23 & +29 18  50.3 &          0.8 &   39.3 &  39.3 \\
NGC 4941        & 2 & 13 04 13.02 & $-$05 33  03.5 &          2.5 &   16.8 &  20.2 \\
NGC 4968        & 2 & 13 07 06.00 & $-$23 40  36.1 &          1.4 &   34.9 &  34.9 \\
NGC 5005        & 2 & 13 10 56.16 & +37 03  32.3 &          1.3 &   86.9 & 181.0 \\
NGC 5033        & 1 & 13 13 27.28 & +36 35  40.4 &          3.4 &   82.2 & 122.7 \\
MCG-3-34-63    & 2 & 13 22 24.58 & $-$16 43  43.1 &          1.8 &  275.3 & 275.3 \\
NGC 5135        & 2 & 13 22 56.51 & $-$29 34  25.6 & $^{\ast}$14.8 &  187.2 & 201.3 \\
NGC 5194=M 51    & 2 & 13 29 52.65 & +47 11  45.7 &          2.9 &  103.6 & 431.7 \\
F13349+2438  & 1 & 13 37 18.73 & +24 23  02.8 &          0.6 &   20.0 &  20.0 \\
NGC 5256=Mrk 266  & 2 & 13 38 17.60 & +48 16  38.4 &          6.6 &  114.1 & 129.7 \\
Mrk 273=UGC 8696  & 2 & 13 44 42.21 & +55 53  13.1 &          0.8 &  134.9 & 145.1 \\
I4329A       & 1 & 13 49 19.23 & $-$30 18  33.7 &          0.6 &   66.8 &  66.8 \\
NGC 5347        & 2 & 13 53 18.04 & +33 29  25.6 &          3.2 &    6.0 &   6.0 \\
Mrk 463        & 2 & 13 56 02.86 & +18 22  19.0 &          0.5 &  381.0 & 381.0 \\
NGC 5506        & 2 & 14 13 14.84 & $-$03 12  27.0 &          0.9 &  339.4 & 339.4 \\
NGC 5548        & 1 & 14 17 59.32 & +25 08  13.5 &          3.0 &   24.7 &  29.1 \\
Mrk 817        & 1 & 14 36 22.07 & +58 47  41.6 &          2.2 &   11.7 &  11.7 \\
F15091-2107  & 1 & 15 11 59.82 & $-$21 19  00.1 &          1.5 &   47.3 &  47.3 \\
NGC 5929        & 2 & 15 26 06.69 & +41 40  21.0 &          9.5 &   85.8 & 110.0 \\
NGC 5953        & 2 & 15 34 32.70 & +15 11  40.9 &          5.7 &   68.4 &  91.2 \\
UGC 9913=ARP220 & s & 15 34 57.26 & +23 30  11.1 &          1.0 &  326.8 & 326.8 \\
MCG-2-40-4     & 2 & 15 48 24.90 & $-$13 45  28.5 &          1.6 &   26.1 &  30.7 \\
F15480-0344  & 2 & 15 50 41.47 & $-$03 53  17.1 &          1.1 &   42.2 &  42.2 \\
Mrk 509        & 1 & 20 44 09.68 & $-$10 43  23.0 &          2.0 &   16.4 &  19.3 \\
UGC 11680=Mrk 897 & 2 & 21 07 45.86 & +03 52  40.5 &          0.1 &   17.3 &  17.3 \\
NGC 7130=I5135  & 2 & 21 48 19.57 & $-$34 57  04.8 &          0.6 &  169.3 & 190.2 \\
NGC 7172        & 2 & 22 02 01.96 & $-$31 52  10.5 &          0.9 &   33.8 &  37.6 \\
F22017+0319  & 2 & 22 04 19.64 & +03 33  48.7 &          8.2 &   12.1 &  18.3 \\
NGC 7314        & 1 & 22 35 46.43 & $-$26 03  01.7 &          5.4 &    7.7 &  33.5 \\
MCG-3-58-7     & 2 & 22 49 37.08 & $-$19 16  25.5 &          2.3 &   12.7 &  12.7 \\
NGC 7469        & 1 & 23 03 15.61 & +08 52  26.3 &          0.2 &  181.0 & 181.0 \\
NGC 7603=Mrk 530  & 1 & 23 18 56.68 & +00 14  37.2 &          0.9 &   24.9 &  24.9 \\
NGC 7674=Mrk 533  & 2 & 23 27 56.69 & +08 46  43.3 &          1.0 &  221.4 & 221.4 \\
NGC 1241        & 2 &      $-$ &      $-$ &          $>$10 &    2.3 &   2.3 \\
CGCG381-051  & 2 &      $-$ &      $-$ &          $>$10 &    2.3 &   2.3 \\
UGC 7064        & 1 &      $-$ &      $-$ & $^{\ast}$$>$20 &    2.3 &   2.3 \\
NGC 4501        & 2 &      $-$ &      $-$ & $^{\ast}$$>$20 &    2.3 &   2.3 \\
MCG-6-30-15    & 1 &      $-$ &      $-$ & $^{\ast}$$>$20 &    2.3 &   2.3 \\
\hline
\end{tabular}
\caption{The NVSS counterparts of Seyferts from the extended 12 $\mu$m sample.}
\label{NVSS.tab}
\end{center}
\end{table*}

\section{Statistical test results}
\label{tests.sec}

We have estimated the power of a statistical test
to reject a null hypothesis by making trials with the observational
data (see Section \ref{compact.sec}).

Table \ref{lPsim.tab} shows the results of 
scaling the 8.4 GHz A--configuration 
luminosity of the Seyfert 1 sub--population 
and repeating two--sample tests.

Table \ref{Dsim.tab} shows the results of 
scaling the radio sizes of the Seyfert 1 sub--population 
and repeating two--sample tests.

\begin{table}
\scriptsize
\begin{center}
\begin{tabular}{|c|r|r|r|r|r|} \hline 
{\bf Scaling}& \multicolumn{5}{c|}{\bf P(null hypothesis)} \\ \cline{2-6}
{\bf Factor} & {\bf Gehan's} & {\bf Gehan's} & {\bf Logrank} & {\bf Peto}
& {\bf Peto \&} \\
  & {\bf (perm.)} & {\bf (hyper.)}&         & {\bf \&Peto}&  {\bf Prentice} \\ \hline 
1/8  &    0.19\% &   0.20\% &   0.60\% &   0.20\% &   0.18\% \\
1/6  &    0.91\% &   0.94\% &   2.54\% &   0.92\% &   0.90\% \\
1/4  &    4.45\% &   4.59\% &   7.84\% &   4.38\% &   4.40\% \\
1/2  &   36.94\% &  37.25\% &  37.56\% &  36.20\% &  36.61\% \\
1    &   70.86\% &  70.87\% &  77.76\% &  71.22\% &  71.48\% \\
2    &   13.96\% &  13.60\% &  26.09\% &  14.43\% &  14.35\% \\
4    &    0.85\% &   0.70\% &   4.14\% &   0.91\% &   0.83\% \\
6    &    0.09\% &   0.06\% &   0.51\% &   0.09\% &   0.07\% \\
8    &    0.03\% &   0.01\% &   0.21\% &   0.03\% &   0.02\% \\
\hline
\end{tabular}
\caption{The probability of the null hypothesis that type 1 and type 2 
Seyferts have 8.4 GHz A--configuration luminosities drawn at random 
from the same parent population varies as scaling 
factors of one--eighth to eight are applied to the type 1 sub--population.
The luminosity of each type 1 Seyfert was multiplied by the scaling
factor and the probability of the null hypothesis was re--calculated. 
The probability of the null hypothesis is less than 5\% for all 
statistical tests with scaling factors less
than or equal to one sixth, or greater than or equal to four. 
}
\label{lPsim.tab}
\end{center}
\end{table}

\begin{table}
\scriptsize
\begin{center}
\begin{tabular}{|c|r|r|r|r|r|} \hline 
{\bf Scaling}& \multicolumn{5}{c|}{\bf P(null hypothesis)} \\ \cline{2-6}
{\bf Factor} & {\bf Gehan's} & {\bf Gehan's} & {\bf Logrank} & {\bf Peto}
& {\bf Peto\&} \\
  & {\bf (perm.)} & {\bf (hyper.)}&         & {\bf \&Peto}&  {\bf Prentice}  \\ \hline 
1/10 &   0.07\%  &   0.06\% &   0.27\%  &   0.16\% &   0.02\% \\
1/8  &   0.12\%  &   0.11\% &   0.68\%  &   0.29\% &   0.17\% \\
1/6  &   0.35\%  &   0.34\% &   1.86\%  &   0.74\% &   0.46\% \\
1/4  &   1.34\%  &   1.34\% &   6.17\%  &   2.51\% &   2.17\% \\
1/2  &  11.37\%  &  11.53\% &  29.06\%  &  16.82\% &  16.20\% \\
1    &  53.00\%  &  53.16\% &  78.93\%  &  59.69\% &  59.51\% \\
2    &  89.34\%  &  89.31\% &  82.46\%  &  87.00\% &  86.97\% \\
4    &  23.53\%  &  22.71\% &  32.29\%  &  24.82\% &  23.77\% \\
6    &   8.65\%  &   7.78\% &  16.82\%  &  10.54\% &   9.32\% \\
8    &   3.07\%  &   2.45\% &   7.88\%  &   4.26\% &   3.28\% \\
10   &   1.42\%  &   1.00\% &   3.95\%  &   2.19\% &   1.46\% \\
\hline
\end{tabular}
\caption{The probability of the null hypothesis, that type 1 and type 2 
Seyferts have 8.4 GHz A--configuration sizes drawn at random 
from the same parent population, varies as scaling 
factors of one tenth to ten are applied to the type 1 sub--population.
The size of each type 1 Seyfert was multiplied by the scaling
factor and the probability of the null hypothesis was re--calculated. 
The probability of the null hypothesis is less than 5\% for all 
statistical tests with scaling factors less
than or equal to one sixth and greater than or equal to 10.
}
\label{Dsim.tab}
\end{center}
\end{table}

\section{Space densities of the extended 12--micron Seyferts}
\label{difflum.sec}

The differential radio luminosity function shown in Figure \ref{Alum.fig} 
has been estimated
using a method based on the V/V$\rm _{max}$ method described by
\scite{Huchra73}.
The density of sources per luminosity interval, $\Phi$(L) is given by,

\begin{equation}
\rm \Phi (L) = \frac{4\pi}{\Omega f \Delta L} \sum_{i=1}^{n} \frac{1}{V_{max}(L)}~~~~~~({\rm Mpc^{-3}mag^{-1}}),
\label{phi-basic}
\end{equation}

where $\Omega$ is the solid angle covered by the sample (7.26 steradians), 
{\it f} is the fraction of the sample for which radio observations  
were obtained (38/47 for Seyfert 1s and 48/60 for Seyfert 2s), 
$\Delta$L = 10$^{0.4}$ WHz$^{-1}$ is the bin width and
V$\rm _{max}$(L) is the maximum accessible volume for each source and
is chosen as  the lower of the two values defined by the radio and 12
$\mu$m flux densities and flux limits.
Individual values of  $\Phi$(L) are given in Table \ref{phi.tab}.

No luminosity upper--limits have been used to construct 
the differential radio luminosity function and therefore  
$\Phi$(L) is likely to be under--estimated at low luminosities. 
However, the Kaplan--Meier estimator 
cannot be used to estimate the shape of the differential luminosity
function accurately because it is not strictly defined when the
distribution of censored data points is not random; the radio 
luminosity upper--limits we have obtained for 12 sources
are grouped towards the lower luminosity end
of the radio luminosity distribution and there is an extremely low 
probability that they were drawn at random from the sample 
(2$\times$10$^{-05}$\% according to the KS test).
Note that non--random censorship does not invalidate two--sample
tests for which accurate estimates of the true distribution functions are
not required. 
The true value of $\Phi$(L) at low
luminosities could be at least a factor of four higher than that given
in Table \ref{phi.tab}; the Kaplan--Meier estimator indicates that
approximately 4 undetected type 1 sources could be found in the
luminosity bin centered at 10$\times$10$^{18.8}$ WHz$^{-1}$, their
effect on $\Phi$(L) depends on the values of V$\rm _{max}$ which are
likely to be defined by their 12$\mu$m flux densities. 

Assigning errors to $\Phi$(L) is not straightforward; in the majority
of cases V$\rm _{max}$(L) values are determined by {\it IRAS} 12 $\mu$m
fluxes which may be dependent on 8.4 GHz VLA A--configuration 
fluxes.
For simplicity, one--sigma fractional uncertainties on $\Phi$(L)
are taken as 1/$\sqrt{N}$, where N is the number of sources in each
luminosity bin.

\begin{table*}
\scriptsize
\begin{center}
\begin{tabular}{|c|c|c|c|c|c|c|c|c|c|} \hline 
$\rm log(L_{rad})$ & \multicolumn{3}{c|}{All Seyferts} & \multicolumn{3}{c|}{Seyfert 1s} & \multicolumn{3}{c|}{Seyfert 2s}\\ 
(bin cent.) & N & $\rm log\Phi$ & $\rm \Delta log\Phi$ & N & $\rm log\Phi$ & $\rm \Delta log\Phi$ & N & $\rm log\Phi$ & $\rm \Delta log\Phi$ \\ \hline 
  18.8 &  2 & $-$4.07 &  0.23 & 1 &  $-$4.41 &   0.30&  1 &  $-$4.35 &  0.30 \\
  19.2 &$-$ & $-$     &   $-$ & $-$ &   $-$  &   $-$ &$-$ &      $-$ &   $-$ \\
  19.6 &  3 & $-$4.52 &  0.20 & 1 &  $-$5.12 &   0.30&  2 &  $-$4.65 &  0.23 \\
  20.0 &  5 & $-$4.29 &  0.16 & 2 &  $-$4.72 &   0.23&  3 &  $-$4.49 &  0.20 \\
  20.4 &  7 & $-$4.70 &  0.14 & 2 &  $-$5.11 &   0.23&  5 &  $-$4.91 &  0.16 \\
  20.8 &  8 & $-$4.97 &  0.13 & 4 &  $-$5.44 &   0.18&  4 &  $-$5.15 &  0.18 \\
  21.2 & 11 & $-$4.73 &  0.11 & 5 &  $-$4.82 &   0.16&  6 &  $-$5.48 &  0.15 \\
  21.6 & 19 & $-$5.14 &  0.09 & 9 &  $-$5.80 &   0.12& 10 &  $-$5.24 &  0.12 \\
  22.0 &  8 & $-$5.44 &  0.13 & 6 &  $-$5.48 &   0.15&  2 &  $-$6.53 &  0.23 \\
  22.4 & 12 & $-$5.50 &  0.11 & 2 &  $-$5.64 &   0.23& 10 &  $-$6.05 &  0.12 \\
  22.8 &  6 & $-$6.33 &  0.15 & 3 &  $-$7.41 &   0.20&  3 &  $-$6.36 &  0.20 \\
  23.2 &  4 & $-$6.51 &  0.18 & 2 &  $-$7.60 &   0.23&  2 &  $-$6.54 &  0.23 \\
  23.6 &$-$ &    $-$  &   $-$ & $-$ &    $-$ &  $-$ &  $-$ &   $-$ &       $-$\\
  24.0 &  1 & $-$7.84 &  0.30 & 1 &  $-$7.84 &   0.30&  $-$ &        $-$ &       $-$\\
\hline
\end{tabular}
\end{center}
\label{phi.tab}
\caption{The space densities of the extended 12 $\mu$m  Seyferts at
8.4~GHz.}
\end{table*}

\section{The radio--infrared luminosity--luminosity plane.}
\label{Malm.sec}

In Section \ref{S8.4-IR.sec} we showed that
evidence of a correlation between  8.4 GHz A--configuration
luminosities and {\it IRAS} 12 $\mu$m luminosities is weak 
when observational limitations are taken into account.
Here we describe the derivation of the limits defined by the Malmquist
bias and the flux limits which are shown in Figure
\ref{12-rad.fig}.

The upper--left portion of Figure \ref{12-rad.fig} is under--populated 
because of the Malmquist bias; 
sources with low 12 $\mu$m luminosities are only
selected from a relatively nearby volume of space in which there is a
low probability of finding sources with high radio luminosities (which
are intrinsically rare).
The dashed line in Figure \ref{12-rad.fig}
represents the limit above which there is less than a 4.6\% probability
of finding a Seyfert and has been estimated
using the following relationship;

\begin{equation}
\rm P(L_{rad}|L_{IR}) = \Phi_{rad}.\frac{1}{\Phi_{IR}}=4.6\%, \label{probphi.eq}
\end{equation}

where P(L$\rm _{rad}|$L$\rm _{IR}$) is the probability of finding a source
with a radio luminosity, L$\rm _{rad}$, given that it has an infrared
luminosity L$\rm _{IR}$, $\rm \Phi_{rad}$ is the space density of
Seyferts per radio magnitude and $\rm \Phi_{IR}$ is the
space density of Seyferts per infrared magnitude
(1/$\rm \Phi_{IR}$ is the characteristic volume sampled in a given 
infrared luminosity interval). 
The values of L$\rm _{rad}$ and L$\rm _{IR}$ which correspond to  $\rm \Phi_{rad}$
and $\rm \Phi_{IR}$ have been estimated from the
differential 12 $\mu$m luminosity function \cite{RMS93} and the
differential 8.4 GHz luminosity function (Fig.
\ref{Alum.fig}c).

The lower--right portion of Figure \ref{12-rad.fig} is
under--populated because of a combination of the infrared and radio
flux limits; a given infrared luminosity defines a maximum distance
at which a source may be detected above the infrared flux
limit and the radio detection threshold defines the minimum 
detectable radio luminosity for sources at that distance.
The dotted line in Figure \ref{12-rad.fig} shows the
minimum detectable radio luminosity at the 
maximum distance allowed by a given infrared luminosity assuming an 
infrared flux limit of 0.22 Jy and
radio detection threshold of 162 $\mu$Jy;
distant sources are unlikely to be found below this limit.
Note that the radio detection threshold varied from source to source 
according to the map quality.

\section{Newly--identified Seyferts in the extended 12--micron sample.}
\label{newSy.sec}

Table \ref{newSy.tab} shows 38 sources which were classified as 
`normal galaxies' or `high--far--infrared' galaxies by Rush et al.
but which are classified as Seyferts in the NED.
Since NED classifications are not always reliable, an accurate
re--definition of the sample will require careful inspection of individual
spectra; where we are aware of them, we have provided 
references to alternative classifications.
Table \ref{newSy.tab} is arranged as follows;
{\it Column 1:} Galaxy name;
{\it Column 2:} NED Seyfert type;
{\it Column 3:} Alternative galaxy type as taken from the following 
sources: 
(1) \scite{Veron97A&A}
(2) \scite{Kinney93} 
(3) \scite{Lancon96}
(4) \scite{Forbes92} 
(5) \scite{Kinney84}
(6) \scite{Calzetti94}
(7) \scite{Taniguchi99}
(8) \scite{Goncalves99}.

Another AGN from the extended 12 $\mu$m sample 
which was classified as a `normal galaxy' by Rush et al. is the
exceptional source NGC 5232.
The optical spectrum of this source is similar to a normal,
early--type galaxy \cite{Drinkwater97}, but it has 
a higher NVSS to {\it IRAS} flux ratio (0.56) than all
92 extended 12 $\mu$m Seyferts for which NVSS data is available.
This source is possibly one of the few examples of a spiral radio galaxy;
the NVSS map appears to show a hundred--kiloparsec--scale jet
which straddles the nucleus of a SA(rs)0/a host galaxy \cite{Condon95}.

\begin{table}
\scriptsize
\begin{center}
\begin{tabular}{|l|c|c|} \hline 
{\bf Galaxy}  & {\bf Type} & {\bf Type} \\  
  & {\bf (NED)} & {\bf (other)} \\ \hline 
MCG-4-2-18     & 1.8 & \\ 
NGC 613        & Sy & Composite (1)  \\      
M+0-7-41       & 2   &  \\        
NGC 1433       & 2 & HII (2)   \\ 
MCG-5-11-6     & 2 &     \\   
F04259-0440    & 2 &  \\  
NGC 1614       & 2 & HII (3)  \\   
NGC 1672       & 2 & Transition (1)   \\  
NGC 1808       & 2 & HII (4)   \\   
NGC 2655       & 2 &    \\         
NGC 2782       & 1 & HII (5)  \\    
NGC 3094       & 2 &     \\   
NGC 3147       & 2 &     \\     
NGC 3367       & Sy &   \\   
NGC 3486       & 2 &     \\      
NGC 3593       & 2 &     \\    
NGC 3735       & 2 &      \\   
NGC 3822       & 2 &    \\      
NGC 3976       & 2 &   \\          
NGC 4699       & Sy &   \\ 
NGC 4725       & Sy &   \\   
NGC 4995       & Sy &   \\         
Markarian 1361 & 2 &     \\ 
NGC 5899       & 2 &     \\   
NGC 5905       & 1 &  Transition (1) \\ 
CGCG 022-021  & 1.9 &     \\  
UGC 9944       & 2 &     \\    
NGC 6217       & 2 & HII (6) \\ 
NGC 6240       & 2 &  \\     
NGC 6552       & 2 &     \\      
IC 5169        & 2 &     \\   
ESO 344-G16    & 1.5 &      \\       
ESO 148-IG02   & 2 & HII (7)  \\  
IC 5298        & 2 & HII (8) \\ 
NGC 7678       & 2 &    \\  
NGC 7733/4     & 2 &     \\ 
MCG+2-60-17    & 2 &      \\   
Markarian 331  & 2 & HII (3)  \\
\hline
\end{tabular}
\end{center}
\caption{Sources classified as galaxies or high--infrared galaxies by
Rush et al. (1993) which are classified as Seyferts in the NED.}
\label{newSy.tab}
\end{table}

\bibliography{12mic}
\bibliographystyle{mnras}

\end{document}